\documentclass[12pt]{article}

\catcode`\@=11
\@addtoreset{equation}{section}

\global\arraycolsep=2pt
\usepackage{amsbsy,amssymb,latexsym,amsfonts,amsmath}
\usepackage{graphicx,color}
\usepackage{ulem}

\allowdisplaybreaks
\newcommand{\bea}{\begin{eqnarray}}
\newcommand{\eea}{\end{eqnarray}}
\newcommand{\bean}{\begin{eqnarray*}}
\newcommand{\eean}{\end{eqnarray*}}

\newcommand{\be}{\begin{equation}}
\newcommand{\ee}{\end{equation}}
\newcommand{\bitm}{\begin{itemize}}
\newcommand{\eitm}{\end{itemize}}
\newcommand{\bmat}{\begin{pmatrix}}
\newcommand{\emat}{\end{pmatrix}}
\newcommand{\bal}{\begin{aligned}}
\newcommand{\eal}{\end{aligned}}
\newcommand{\ba}{\begin{align}}
\newcommand{\ea}{\end{align}}
\newcommand{\bat}[1]{\begin{alignat}{#1}}
\newcommand{\eat}{\end{alignat}}
\newcommand{\bse}{\begin{subequations}}
\newcommand{\ese}{\end{subequations}}

\newcommand{\re}{\mathbb{R}}

\newcommand{\Tr}{\text{Tr}}
\newcommand{\str}{\textrm{STr}}

\renewcommand{\footnote}[1]{\textcolor{red}{\footnotemark}\footnotetext{\,#1}}

\def\rmd{{\rm d}}

\newcommand{\Slash}[1]{{\ooalign{\hfil#1\hfil\crcr\raise.167ex\hbox{/}}}}

\begin{document}
\begin{titlepage}
\begin{flushright}
\end{flushright}

\vspace*{-1cm}

\centerline{\Large \bf Super-$W_\infty$ Asymptotic Symmetry}
\vskip0.35cm
\centerline{\Large \bf of}
\vskip0.35cm
\centerline{\Large \bf Higher-Spin AdS$_3$ Supergravity}
\vspace*{1cm}
\centerline{ Marc Henneaux $^{a,b}$, Gustavo Lucena G\' omez $^{a}$, Jaesung Park $^c$, Soo-Jong Rey $^{c,d}$}
\vspace*{1cm}
\centerline{$^a$ \sl Universit\'e Libre de Bruxelles (ULB) \& International Solvay Institute (ISI)}
\vspace*{.15cm}
\centerline{\sl Campus Plaine CP231, B-1050 \rm BELGIUM}
\vspace*{.15cm}
\centerline{${}^b$ \sl Centro de Estudios Cient\'{\i}ficos (CECS), Casilla 1469, Valdivia \rm CHILE}
\vspace*{0.15cm}
\centerline{${}^c$ \sl School of Physics and Astronomy, Seoul National University, Seoul 151-747 \rm KOREA}
\vspace*{0.15cm}
\centerline{${}^d$ \sl The Simons Center for Geometry and Physics, NY 11790 \rm USA}
\vspace{1.75cm}
\centerline{ABSTRACT}
\vskip0.75cm
\noindent
We consider $(2+1)$-dimensional $(N, M)$-extended higher-spin anti-de Sitter supergravity and study its asymptotic symmetries. The theory is described by the Chern-Simons action based on the real, infinite-dimensional higher-spin superalgebra $\textrm{shs}^\textrm{E}(N|2,\mathbb{R})\oplus\textrm{shs}^\textrm{E}(M|2,\mathbb{R})$. We specify consistent boundary conditions on the higher-spin super-gauge connection corresponding to asymptotically anti-de Sitter spacetimes. We then determine the residual gauge transformations that preserve these asymptotic conditions and compute their Poisson bracket algebra. We find that the asymptotic symmetry is enhanced from the higher- spin superalgebra to some $(N,M)$-extended super-$\textrm{W}_\infty$ nonlinear superalgebra. The latter has the same classical central charge as pure Einstein gravity. Special attention is paid to the $(1,1)$ case.  Truncation to the bosonic sector yields the previously found $\textrm{W}_\infty$ algebra, while truncation to the $\textrm{osp}(N|2, \mathbb{R})$ sector reproduces the $N$-extended  superconformal algebra (in its nonlinear version for $N>2$). We discuss string theory realization of these higher-spin anti-de Sitter supergravity theories as well as relations to previous treatments of super-$\textrm{W}_\infty$ in the literature.

\end{titlepage}
\setcounter{page}{2}

\newpage

\section{Introduction}

Recently, two of the authors \cite{Henneaux:2010xg} identified the asymptotic symmetry algebra of the $(2+1)$-dimensional higher-spin gravity theory of \cite{Vasiliev:1986qx,Blencowe:1988gj} (see also \cite{Campoleoni:2010zq,Gaberdiel:2011wb,Campoleoni:2011hg}). The theory is most compactly given in a framelike formulation, described by the Chern-Simons gauge theory whose gauge algebra is the infinite-dimensional higher-spin algebra $\textrm{hs}(2,\mathbb{R}) \oplus\textrm{hs}(2,\mathbb{R})$ \textcolor{black}{with Chern-Simons levels $+k, -k$.} The ground-state is the three-dimensional anti-de Sitter spacetime (AdS$_3$), whose isometry algebra  coincides with the finite-dimensional gauge subalgebra \textcolor{black}{$\textrm{sl}(2, \mathbb{R}) \oplus \textrm{sl}(2, \mathbb{R})$}. \textcolor{black}{Perturbing the fields around the ground state deforms the space-time but one considers configurations that are still asymptotic to  AdS$_3$ at spacelike infinity.} 

In  \cite{Henneaux:2010xg}, we have shown that the asymptotic symmetry algebra, viz. the spacetime symmetry algebra at spacelike infinity, is given by  the algebra $\textrm{W}_\infty \oplus \textrm{W}_\infty$ with a classical central charge
\bea
c  =  6k  = {3 \ell \over 2 G_N},
\eea
whose value coincides with that of the pure gravity and the (extended) supergravity theories  \cite{Brown:1986nw,Banados:1998pi,Henneaux:1999ib}. Here, the algebra $\textrm{W}_\infty$ is a nonlinear infinite-dimensional version of the Zamolodchikov $\textrm{W}_N$ algebra \cite{Zamolodchikov:1985wn,Bouwknegt:1992wg}, described most easily in terms of the Drinfeld-Sokolov (DS) Hamiltonian reduction  \cite{Drinfeld:1984qv}: as shown in \cite{Coussaert:1995zp}, the boundary conditions that define asymptotically anti-de Sitter spacetimes in pure $(2+1)$-dimensional anti-de Sitter gravity in fact implement the Hamiltonian reduction from the $\textrm{sl}(2,\mathbb{R})$-WZNW model to the Liouville theory, along the lines of  \cite{Forgacs:1989ac,Balog:1990mu}.  The boundary conditions of \cite{Henneaux:2010xg} extend the construction to the higher-spin algebra $\textrm{hs}(2, \mathbb{R})$, which actually provides an alternative scheme of extracting the structure coefficients of the $\textrm{W}_\infty$ algebra \cite{Gaberdiel:2011wb,Campoleoni:2011hg}.

\textcolor{black}{ The asymptotic symmetry algebra identified in \cite{Henneaux:2010xg} revealed several intriguing facts. One is that the  asymptotic symmetry algebra $\textrm{W}_\infty \oplus \textrm{W}_\infty$ is nonlinear and hence not a true Lie algebra. This is in stark contrast with the situation of the spin-2 Einstein gravity, where the asymptotic symmetry algebra is the Virasoro algebra $\textrm{Vir} \oplus \textrm{Vir}$, which is of course a Lie algebra. Another is that the classical central charge of this theory is the same as that of the Einstein gravity. Apparently, higher-spin currents do not contribute to the degrees of freedom.  Both these features were actually already encountered previously in extensions of pure gravity involving a finite number of fields \cite{Henneaux:1999ib} but here, they take a more spectacular manifestation since there is an infinite number of fields besides the spin-2 one.} \textcolor{black}{One hopes to gain a better understanding of this rigidity of the central charge by endowing more symmetry structures. }

In this work, in order to further investigate these properties,  we expand the analysis of  \cite{Henneaux:2010xg} to extended super-higher-spin gravity theories in $2+1$ dimensions, described by the supersymmetric extensions $\textrm{shs}^\textrm{E}(N|2,\mathbb{R}) \oplus \, \textrm{shs}^\textrm{E}(M|2,\mathbb{R})$ of the bosonic higher- spin algebra hs$(2,\re)$. Here, $M$ and $N$ are positive integers labelling the extended supersymmetry.

 \textcolor{black}{It is actually well known that there exist seven different classes of extended supersymmetry \textcolor{black}{AdS} algebras in \textcolor{black}{$2+1$} dimensions (see section 2.1 below), and seven corresponding classes of superconformal algebras with quadratic deformations. Each of these classes leads to a different \textcolor{black}{ higher-spin gauge superalgebra}.  We cover here explicitly the higher-spin extension $\textrm{shs}^\textrm{E}(N \vert 2,\mathbb{R})$ of the $\textrm{osp}(N \vert 2, \re)$ \textcolor{black}{AdS$_3$} superalgebra but, as argued below, \textcolor{black}{ higher-spin extensions of other AdS$_3$ superalgebras} can be treated similarly.} 

\textcolor{black}{Another motivation is that}  \textcolor{black}{higher-spin AdS$_3$ supergravity} opens yet another door to the exploration of the AdS/CFT correspondence -- an arena in which the previous work \cite{Henneaux:2010xg} triggered subsequent investigations. Pursuing ideas made in that work, we first discuss possible string theory realizations of the higher-spin AdS$_3$ supergravity theories from stretched horizon of near-horizon geometry of macroscopic heterotic or Type II strings. It turns out that different compactifications of the transverse directions lead to different types of higher-spin AdS$_3$ supergravity theories whose higher-spin gauge superalgebras are built upon the different finite-dimensional superalgebras listed in Table 1. One also would like to identify the holographic dual spacetime CFTs. Like the bosonic counterpart, the higher-spin AdS$_3$ supergravity theory has two coupling parameters $G_N, \ell$, \textcolor{black}{so} the holographic dual spacetime CFTs must also have two parameters. The naive extension of the bosonic proposal is that the dual spacetime CFTs are supercoset CFTs with two tunable coupling parameters. On the other hand, one generally expects that extended supersymmetries put severe constraints on possible supercoset classes \textcolor{black}{and the CFT parameters}. 
 
Our paper is organized as follows. In section \ref{algebra}, we review the extended super-higher spin algebra $\textrm{shs}^\textrm{E}(N|2,\mathbb{R})\oplus\textrm{shs}^\textrm{E}(M|2,\mathbb{R})$  through its explicit super-oscillator construction. For definiteness, we consider the undeformed case first.  We show how it contains $\textrm{osp}(N|2,\mathbb{R})\oplus\textrm{osp}(M|2,\mathbb{R})$ and investigate some of its other bosonic subalgebras.  We also provide an alternative description of the  unextended case of  $\mathcal{N}=(1,1)$.
In section \ref{SuperCS}, taking $\textrm{shs}(N|2,\mathbb{R})\oplus \textrm{shs}(M|2,\mathbb{R})$ as the gauge \textcolor{black}{superalgebra}, we present the formulation of the super-higher spin gravity theory as a Chern-Simons theory. In section \ref{AsymptoticSymmetries}, we study the asymptotic supersymmetry algebra and show in section \ref{NonlinearW} that it is given by a nonlinear \textcolor{black}{super-$W_\infty$} algebra. At this stage, it  deepens our understanding of higher-spin dynamics to underline the details of how the higher-spin superalgebra is enhanced to the \textcolor{black}{super-W$_\infty$} superalgebra. In section \ref{wedge}, we thus analyze the embedding of the higher-spin superalgebra in the \textcolor{black}{super-W$_\infty$} algebra and show how it can be identified with the wedge subalgebra of the latter, whose definition is recalled there. In section \ref{lambdaDeformed}, we briefly recall what are the $\lambda$-deformations of the higher-spin superalgebras and discuss how extended supersymmetry puts severe constraints on them.  In section \ref{OtherClasses},  we consider
higher-spin superalgebras extending finite-dimensional superalgebras other  than $\textrm{osp}(N|2, \mathbb{R})$, as classified in section 1 and explain why the procedure also goes through in those cases.
In section 9, we discuss candidate superconformal field theories as holographic dual of these higher-spin \textcolor{black}{supergravity} theories.  We also gather further points of discussion including implications to tensionless string theory and quantization of higher-spin \textcolor{black}{supergravity} theory.  The last section is devoted to concluding remarks. In four appendices, we collect our conventions, the matrix realization of the osp$(N|2, \mathbb{R})$ superalgebra, useful relations for the$N=1$ supersymmetric higher-spin superalgebra and the explicit oscillator construction of the higher-spin shs$(N \vert 1,1)$ superalgebra based on the algebra su$(N \vert 1,1)$.

\section{The $\textrm{shs}^\textrm{E}(N|2, \mathbb{R})\oplus\textrm{shs}^\textrm{E}(M|2,\mathbb{R})$ \textcolor{black}{superalgebras}  }
\label{algebra}

\subsection{Extended supergravities and higher spins in $2+1$ dimensions}
As is well-known, the isometry algebra of the three-dimensional anti-de Sitter spacetime (AdS$_3$) is the Lie algebra $\textrm{so}(2,2) \simeq \textrm{sl}(2, \mathbb{R}) \oplus \textrm{sl}(2, \mathbb{R})$.   It has been shown in \cite{Achucarro:1987vz,Witten:1988hc} that three-dimensional Einstein gravity with a negative cosmological constant (AdS$_3$ gravity) can be reformulated as a Chern-Simons gauge theory whose gauge connection take values in the isometry algebra $\textrm{sl}(2, \mathbb{R})_{+k} \oplus \textrm{sl}(2, \mathbb{R})_{-k}$. Here $\pm k$ denotes the Chern-Simons levels of the chiral and  anti-chiral sectors and is related to the gravitational coupling constant through formula (\ref{RelationGk}) below. 

The reformulation can be generalized to ${\cal N} = (N,M)$-extended AdS$_3$ supergravity \cite{Achucarro:1987vz}, where $M, N$ refer to the supersymmetry of the two gauge factors.  Recall the isomorphism
\bea
\textrm{sl}(2, \mathbb{R}) \simeq \textrm{sp}(2,\mathbb{R}) \simeq \textrm{so}(2,1) \simeq \textrm{su}(1,1).
\label{isomorphism}
\eea
Then, AdS$_3$ supergravity theories are obtainable by taking an appropriate superalgebra containing (\ref{isomorphism}) as a bosonic subalgebra. 

For example,  ${\cal N}=(1,1)$ AdS$_3$ supergravity can be reformulated as a Chern-Simons super-gauge theory whose gauge super-connection takes values in the Lie superalgebra $$\textrm{osp}(1|2,  \mathbb{R})_{+k} \oplus \textrm{osp}(1|2, \mathbb{R})_{-k}.$$  Likewise, ${\cal N}= (N, M)$-extended supergravity is based on a super-connection taking values in the Lie superalgebra\footnote{In what follows, we shall omit the Chern-Simons level specification in the gauge superalgebras. They can be reconstructed from the context.} $$\textrm{osp}(N|2, \mathbb{R})_{+k} \oplus \textrm{osp}(M| 2, \mathbb{R})_{-k}.$$ 

In all these cases, either chiral copy contains $\textrm{sp}(2, \mathbb{R})$ as a bosonic subalgebra. The bosonic subalgebra of $\textrm{osp}(N| 2, \mathbb{R})$ is actually of the form $\textrm{sp}(2,\mathbb{R}) \oplus \mathcal{G}$,  where $ \mathcal{G}$ is here $\textrm{so}(N)$. The fermonic generators transform as spinors of $\textrm{sp}(2,\mathbb{R})$ and vectors of so($N$).  

More generally, one can take the gauge superalgebra to be a direct sum of two simple superalgebras ${\cal A}_L, {\cal A}_R$:
\bea
{\cal A}_L \oplus {\cal A}_R,
\eea
with the conditions that (i) each superalgebra contains any of (\ref{isomorphism}) as a bosonic subalgebra; and (ii) the fermionic generators transform in the $\bf{2}$ of (\ref{isomorphism}). It has been shown \cite{Kac,Nahm:1977tg,Gunaydin:1986fe} that these conditions are satisfied in only seven classes, which are listed in Table 1. 
Thus, the most general $(N,M)$-extended AdS$_3$ supergravity can be defined as the Chern-Simons gauge theory whose gauge super-connections in the chiral and anti-chiral sectors take values in any two of the seven Lie superalgebras of Table 1.

\vspace{.5cm}

\begin{table}
\label{Table1}
\begin{centering}
\begin{tabular}{|c|c|c|c|}
  \hline
  $\mathcal{A}$ & $\mathcal{G}$  & $\rho$ & $D$ \\
  \hline
  osp$(N|2, \mathbb{R})$ & so$(N)$ & ${\bf N}$ & $\frac{N(N-1)}{2}$ \\
  su$(1,1 \vert N)$ ($N \not=2$) & su$(N)$  $\oplus$ u$(1)$  &${\bf N} + \overline{\bf N}$ & $N^2$ \\
   su$(1,1 \vert 2)$ $/$ u$(1)$ & su$(2)$ &${\bf 2} + \overline{\bf 2}$ & $3$ \\
    osp$(4^* \vert 2M)$  & su$(2)$ $\oplus$ usp$(2M)$ &$({\bf 2M},{\bf 2})$  & $M(2M+1) + 3$ \\
D$^1 (2,1; \alpha)$  & su$(2)$ $\oplus$ su$(2)$ &$({\bf 2},{\bf 2})$  & $6$ \\
G$(3)$  & G$_2$  &${\bf 7}$  & $14$ \\
F$(4)$  & spin$(7)$  &${\bf 8}_\textrm{s}$  & $21$ \\
  \hline
\end{tabular}
\caption{\footnotesize{Superalgebras of extended anti-de Sitter supegravities in $2+1$ dimensions. Here, ${\mathcal A}$ is the (extended) superalgebra, ${\mathcal G}$ is the internal subalgebra, $\rho$ is the representation of ${\mathcal G}$  in which the spinors transform, and $D$ is the dimension of ${\mathcal G}$.} The first four superalgebras belong to the osp$(m,2n)$  and spl$(m,n)$ infinite families, while the last three are the ``exceptional" Lie superalgebras.}
\end{centering}
\end{table}

Just as extended AdS$_3$ supergravity can be formulated as a Chern-Simons super-gauge theory,  consistent higher-spin AdS$_3$ supergravity theories can also be formulated as Chern-Simons super-gauge theories  \cite{Blencowe:1988gj}.  This time, however, the  gauge superalgebras ${\cal A}_L, {\cal A}_R$ are infinite-dimensional. Since the standard AdS$_3$ supergravity ought to be a consistent truncation of these theories, it must be that these infinite-dimensional gauge superalgebras
contain the simple superalgebras in Table 1 as subalgebras. In other words, the higher-spin superalgebras are infinite-dimensional extensions of these simple superalgebras.  Note that the Table 1 lists the superalgebras 
relevant for `pure' supergravity. Once matter supermultiplets are coupled, the 
total number of supersymmetries would be limited maximally to 32. These 
couplings will be discussed in sections 7 and 9.

We shall mostly concentrate on the osp$(N|2, \mathbb{R})$ class, since this is the class that encompasses uniformly all extended supersymmetries on each chiral sector. In section \ref{OtherClasses}, however, we shall touch on the other six classes, and demonstrate that the construction of the corresponding higher-spin AdS$_3$ supergravities is a straightforward generalization of the foregoing construction in the osp$(N |2, \mathbb{R})$ case.
 
We first need an infinite-dimensional extension of osp$(N|2, \mathbb{R}) \oplus \textrm{osp}(M|2, \mathbb{R})$ to a suitable higher-spin superalgebra. Fortunately, the construction of the relevant superalgebras were worked out already in \cite{Vasiliev:1986qx,Blencowe:1988gj}.  These superalgebras are denoted $\textrm{shs}^\textrm{E}(N|2,\mathbb{R})\oplus\textrm{shs}^\textrm{E}(M|2,\mathbb{R})$\footnote{The two simple algebras describe each chiral sector and can be analyzed separately. From now on we shall focus on the chiral part.}. In a nutshell, the higher-spin superalgebra so constructed corresponds to the universal enveloping superalgebra of underlying finite-dimensional sub-superalgebras osp$(N |2, \mathbb{R}) \oplus$ osp$(M | 2, \mathbb{R})$ quotiented by certain ideals. This fits also with the requirement that the standard AdS$_3$ supergravity be a consistent truncation of the higher-spin AdS$_3$ supergravity. 

In this section, we explain the higher-spin superalgebra shs$^E(N| 2, \mathbb{R})$ and its simplest realization  in terms of ``super-oscillators". In this realization, the minimal $N=1$ case is special since it admits another equivalent formulation with a smaller number of oscillators. We shall mention this aspect along the way as we discuss the general $N$ cases. 

\subsection{Polynomial realization of $\textrm{shs}^\textrm{E}(N|2,\mathbb{R})$}
\label{erealization}
In this section, we realize the Lie super-algebra shs$^\textrm{E}(N | 2, \mathbb{R})$ in terms of super-oscillator  polynomials.

\subsubsection{General $N$}

Consider the following $N+2$ Grassmann variables: two commuting ones, $q_\alpha$ ($\alpha = 1,2$), together with $N$ anticommuting ones, $\psi_i$ ($i=1,\dots,N$). Adapting to the terminology used in the literature, we refer to the index $i$ as the `color' index. As such,
\bat{2}
q_\alpha q_\beta &= q_\beta q_\alpha {}&{} &\qquad\forall\; \alpha,\beta = 1,2 \\
\psi_i\psi_j &= - \psi_j\psi_i {}&{} &\qquad\forall\; i,j = 1,\dots , N \\
q_\alpha\psi_i &= \psi_i q_\alpha {}&{} &\qquad\forall\; \alpha = 1,2\quad \&\quad i = 1,\dots , N .
\end{alignat}
These variables are all taken to be real-valued, $q_\alpha^*= q_\alpha$, $\psi_i^* = \psi_i$.    We construct polynomials in these $N+2$ variables, with coefficients that can be themselves commuting or anticommuting, i.e.,  that belong also to a different Grassmann algebra ${\mathcal G}$.  Thus, we formally consider the (graded) tensor product ${\mathcal A} = {\mathcal G} \otimes {\mathcal P}$ of the polynomial algebra ${\mathcal P}$ in $q_\alpha$, $\psi_i$ with the Grassmann algebra ${\mathcal G}$. The sign in the commutation relations for the multiplication of elements in the graded tensor product is dictated by the total grading, so that odd elements of ${\mathcal G}$ and ${\mathcal P}$ anticommute.  The Grassmann parity used below will always be the total grading. A complex conjugation is assumed to be defined in ${\mathcal G}$, and can be extended to ${\mathcal A}$ taking into account that $q_\alpha$ and  $\psi_i$ are real. We systematically use the convention $(ab)^*= b^* a^*$.

Let  ${\mathcal A}^E$ be the subalgebra of Grassmann-even polynomials in $q_\alpha$, $\psi_i$ containing only monomials of even degree and no constant term.  Thus, a general element of ${\mathcal A}^E$ reads
\be
\bal
f &=  f^{\alpha\beta} q_\alpha q_\beta + f^{\alpha, i} q_\alpha \psi_i  +  f^{ij}\psi_i\psi_j  \\
& + f^{\alpha\beta \gamma \delta}q_\alpha q_\beta q_\gamma q_\delta+ f^{\alpha\beta \gamma,i}q_\alpha q_\beta q_\gamma \psi_i  + f^{\alpha\beta, ij}q_\alpha q_\beta \psi_i\psi_j +  \dots \\
& + f^{\alpha\beta \gamma \delta\epsilon\eta }q_\alpha q_\beta q_\gamma q_\delta q_\epsilon q_\eta+ \dots \\
& + \dots ,
\eal
\label{poly}
\ee
where terms of arbitrarily high power are allowed.  The coefficients in this expansion are completely symmetric (respectively, antisymmetric) in the Greek (respectively, Latin) indices.  They are commuting (respectively, anticommuting) whenever they multiply an even (respectively, odd) number of $\psi$'s. When we formulate higher-spin AdS$_3$ supergravity as a Chern-Simons super-gauge theory, the gauge super-connection will be taken to be of the form (\ref{poly}). The coefficients in the expansion will then be identified with commuting or anticommuting spacetime fields of higher spins.

A $\star$-product  is defined on  ${\mathcal A}$ as follows:
\bea
\label{estarprod}
(f\star g) (z'')\equiv \exp\left(i\,\epsilon_{\alpha\beta}\frac{\partial}{\partial q_\alpha}\frac{\partial}{\partial q'_\beta} + \delta_{ij}\frac{\overleftarrow{\partial}}{\partial\psi_i}\frac{\overrightarrow{\partial}}{\partial\psi'_j}\right)f(z)g(z')\,{\vphantom{\frac{1}{2}}\vline}_{\,z = z' = z''}\, ,
\label{starproduct}
\eea
where $f(z)\equiv f(q_\alpha,\psi_i)$ and so on. In this expression, $f(z)g(z')$ is the standard Grassmann product. The operation (\ref{estarprod}) is called the $\star$-product.
Left and right derivatives with respect to the anticommuting variables are defined by
\be
\bal
\label{leftderivative}
\delta f &= \delta \psi_i \frac{\overrightarrow{\partial} f}{\partial\psi_i} \\
\delta f &= \frac{\overleftarrow{\partial} f}{\partial\psi_i} \delta \psi_i .
\eal
\ee
The epsilon symbol is explicitly taken to be
\be
(\epsilon^{\alpha\beta}) \equiv (\epsilon_{\alpha\beta}) \equiv \bmat \ 0 & \  1 \\ -1 &  \ 0 \emat ,\qquad \alpha,\,\beta \in \{1,2\}.
\ee
More information about our conventions is gathered in appendix \ref{A}.

The above $\star$-product is well known to be associative. However, it does not preserve the reality condition, in the sense that $f\star g$ is not real even if $f$ and $g$ are so.  On the other hand, one can check that if $f$ and $g$ are both real elements of ${\mathcal A}^E$, or both pure imaginary elements of ${\mathcal A}^E$,  of respective order $2n$ and $2m$, then the homogenous polynomials appearing in the expansion of $f \star g$,
\be f \star g = \sum_{j=0}^{m+n}  h_{2(m+n-j)} \, ,\label{expansionfstarg}
\ee
are alternatively real and imaginary.  More precisely,  the homogeneous polynomial $h_{2(m+n-j)}$  of degree $2(m+n-j)$ in $q_\alpha$, $\psi_i$ is:
\begin{itemize}
\item real and symmetric for the exchange of $f$ and $g$ when $j$ is even;
\item imaginary and antisymmetric for the exchange of $f$ and $g$ when $j$ is odd.
\end{itemize}
We then define the $\star$-commutator (also called ``$\star$-bracket"),
\be
\label{Liebracket}
[f,g]_\star \equiv f\star g - g\star f \, ,
\ee
which fulfills the Jacobi identity since the $\star$-product is associative.   From what we have just seen,   $[f,g]_\star$ is pure imaginary whenever $f$ and $g$ are both real or both pure imaginary.  

The Lie superalgebra $\textrm{shs}^\textrm{E}(N|2,\mathbb{R})$ is the real subspace of pure imaginary elements of ${\mathcal A}^E$ equipped with the $\star$-bracket\footnote{One could equivalently insert a factor of $i$ in the definition of the $\star$-bracket, which would no longer coincide with the star commutator, and define $\textrm{shs}^\textrm{E}(N|2,\mathbb{R})$ as the subspace of real polynomials equipped with that alternative bracket.   Either convention has its own advantages.}.  A general element of $\textrm{shs}^\textrm{E}(N|2,\mathbb{R})$ is thus of the above form
\be
\bal
f &=  f^{\alpha\beta} q_\alpha q_\beta + f^{\alpha, i} q_\alpha \psi_i  +  f^{ij}\psi_i\psi_j  \\
& + f^{\alpha\beta \gamma \delta}q_\alpha q_\beta q_\gamma q_\delta+ f^{\alpha\beta \gamma,i}q_\alpha q_\beta q_\gamma \psi_i  + f^{\alpha\beta, ij}q_\alpha q_\beta \psi_i\psi_j +  \dots \\
& + f^{\alpha\beta \gamma \delta\epsilon\eta }q_\alpha q_\beta q_\gamma q_\delta q_\epsilon q_\eta+ \dots \\
& + \dots ,
\eal
\ee
but the coefficients are further restricted so as to make $f$  imaginary.  So, for instance, the coefficient $f^{\alpha\beta}$ is imaginary while $f^{\alpha, i}$ and $f^{ij}$ are real.

One can rewrite alternatively (\ref{Liebracket}) as
\be
\label{superLiecomm}
\bal
~[f,g]_\star (z'') &= \left(2 i \sin\left(\epsilon_{\alpha\beta}\frac{\partial}{\partial q_\alpha}\frac{\partial}{\partial q'_\beta}\right)\cosh\left(\delta_{ij}\frac{\overleftarrow{\partial}}{\partial\psi_i}\frac{\overrightarrow{\partial}}{\partial\psi'_j}\right)\right. \\
&\hspace*{0.8pt}+\left. 2\cos\left(\epsilon_{\alpha\beta}\frac{\partial}{\partial q_\alpha}\frac{\partial}{\partial q'_\beta}\right)\hspace*{1.3pt}\sinh\left(\delta_{ij}\frac{\overleftarrow{\partial}}{\partial\psi_i}\frac{\overrightarrow{\partial}}{\partial\psi'_j}\right)\right)f(z)g(z')\,{\vphantom{\frac{1}{2}}\vline}_{\,z = z' = z''} .
\eal
\ee

It should be stressed that the polynomial $[f,g]_\star$ starts at highest polynomial degree $2(n+m-1)$.  Note also  that the lowest polynomial degree term in the expansion (\ref{expansionfstarg}) is $h_{2(\vert m - n \vert)}$ so that there is a term of degree zero in (\ref{expansionfstarg}) only if $n=m$, in which case $j=2m$ is even. This implies that the term of degree zero (when present) is symmetric for the exchange of $f$ with $g$ and in particular that the constant term (when present in $f \star g$) drops from the $\star$-commutator so that $[f,g]_\star$ has indeed no constant term and belongs to $\textrm{shs}^\textrm{E}(N|2,\mathbb{R})$.

\subsubsection{Supertrace and scalar product}

The supertrace of a polynomial in the $q$'s and the $\psi$'s is defined by its component of degree zero:
\be \str f (q,\psi) = 8 f( 0) . \ee
The normalization is chosen to match standard conventions in the normalization of the action below. Thus, elements in $\textrm{shs}^\textrm{E}(N|2,\mathbb{R})$ all have zero supertrace. 

Although $ \str f =0$ $\forall f \in \textrm{shs}^\textrm{E}(N|2,\mathbb{R})$, it turns out that $\str (f \star g) $ may differ from zero even if $f,g \in \textrm{shs}^\textrm{E}(N|2,\mathbb{R})$.  One thus defines a scalar product on $\textrm{shs}^\textrm{E}(N|2,\mathbb{R})$ by
\be
\label{scalprod}
(f,g) \equiv \str (f\star g).
\ee
The scalar product is evidently  bilinear, real and symmetric (given our discussions in the previous subsection). Using the symmetry together with the associativity of the $\star$-product, we further conclude that it is also invariant:
\be
([f,g]_\star,h) = (f,[g,h]_\star).
\ee
In addition, it is non-degenerate.  It is non-zero only when $f$ and $g$ have same degree in both the $\psi_i$'s and the $q_\alpha$'s. It is this scalar product that will be used to define the Chern-Simons action below.

\subsubsection{Basis}
A basis of $\textrm{shs}^\textrm{E}(N,2|\mathbb{R})$ is given by the monomials
\be
\label{basis0}
X_{p,q;\, i_1,i_2, \cdots, i_N} \equiv  \frac{i^{\lfloor \frac{K+1}{2}\rfloor}}{2\, i \, p! \,q!}q_{1}^p q_{2}^q \psi_1^{i_1}\dots\psi_N^{i_N} ,
\ee
where $p,q \in\mathbb{N}$ and $ i_k \in \{0,1\}$. The degree of $X_{p,q;\, i_1,i_2, \cdots, i_N}$, which  is $p+q+ K$, must be even and positive, where $K = \sum_{k=1}^N i_k$ is the degree in the $\psi$'s. The power of $i$ has been inserted in such a way that the elements of even Grassman parity are imaginary, while those of odd Grassman parity are real.

With this choice, a general element of $\textrm{shs}^\textrm{E}(N|2,\mathbb{R})$ is of the form
\be
\sum \mu^{p,q;\, i_1,i_2, \cdots, i_N} X_{p,q;\, i_1,i_2, \cdots, i_N}
\ee
where the coefficients $  \mu^{p,q;\, i_1,i_2, \cdots, i_N}$ are real and of Grassman parity $(-1)^K= (-1)^{p+q}$.

\subsubsection{$\textrm{osp}(N|2,\mathbb{R})$ sub-superalgebra}
The subspace of quadratic polynomials is a subalgebra isomorphic to $\textrm{osp}(N|2,\mathbb{R})$,  as is known from the familiar oscillator realization of $\textrm{osp}(N|2,\mathbb{R})$ \cite{Gunaydin:1986fe}.  Renormalizing and relabeling\footnote{Note that we have changed the letter $X$ to $Y$ for the generators with no $\psi$'s since these differ from the corresponding $X$'s by a factor.} the quadratic basis elements as
\bea
 Y_{\alpha \beta} = - \frac{i}{2} q_\alpha q_\beta, \; \; \; X_{\alpha i } = \frac{1}{2} q_{\alpha}\psi_i, \; \; \; X_{ij} = \frac{1}{2} \psi_i \psi_j \ , 
\label{basiselement}
\eea
one finds that the non-zero Lie superbrackets read explicitly
\be
\bal
~[Y_{\alpha\beta},Y_{\gamma\delta}]_\star &= \epsilon_{\alpha\gamma}Y_{\beta\delta} + \epsilon_{\alpha\delta}Y_{\beta\gamma} + \epsilon_{\beta\gamma}Y_{\alpha\delta} + \epsilon_{\beta\delta}Y_{\alpha\gamma} \\
[X_{\alpha i},Y_{\beta\gamma}]_\star &= \epsilon_{\alpha\beta}X_{\gamma i} + \epsilon_{\alpha\gamma}X_{\beta i} \\
\{X_{\alpha i},X_{\beta j}\}_\star &= i \left(\epsilon_{\alpha\beta}X_{ij} - \delta_{ij}Y_{\alpha\beta} \right)
\\
[X_{ij},X_{\alpha k}]_\star &= \delta_{jk}X_{\alpha i} - \delta_{ik}X_{\alpha j} \\
[X_{ij},X_{kl}]_\star &= \delta_{il}X_{jk} + \delta_{jk}X_{il} - \delta_{ik}X_{jl} - \delta_{jl}X_{ik}.
\eal
\label{osp-commutator}
\ee

Hence, one goes from $\textrm{shs}^\textrm{E}(N|2,\mathbb{R})$ to $\textrm{osp}(N|2,\mathbb{R})$ by restricting the  $\star$-algebra of polynomials of even degree in the $q$'s and the $\psi$'s to the $\star$-subalgebra of polynomials of  degree two. Conversely, one goes from $\textrm{osp}(N|2,\mathbb{R})$ to $\textrm{shs}^\textrm{E}(N|2,\mathbb{R})$ by relaxing the condition that the polynomials should be quadratic, i.e., by allowing arbitrary (pure imaginary) polynomials of even degree modulo zero-degree term.

The $\textrm{osp}(N|2,\mathbb{R})$ subsuperalgebra can also be realized in terms of matrices. In a matrix representation where imaginary elements are represented by anti-hermitian matrices for an appropriate (indefinite) hermitian product, the $Y_{\alpha \beta}$ are elements of $\textrm{su}(1,1) \simeq \textrm{sl}(2, \mathbb{R})$. Though we shall primarily use the oscillator polynomial realization, for comparison and completeness we collect the relevant results on the matrix representation in Appendix B.

As already mentioned, the infinite-dimensional higher-spin superalgebra corresponds to the universal enveloping superalgebra of the underlying finite-dimensional sub-superalgebra quotientized by certain ideals. The latter being generated by quadratic polynomials ${\cal A}^{(2)}$, this means that the polynomials of ${\cal A}^E$ can be reexpressed as polynomials of the generators of the finite-dimensional sub-superalgebra ${\cal A}^{(2)}$. This point of view is useful when considering the construction of the other higher spin superalgebras (see section \ref{OtherClasses}).

\subsubsection{ $\textrm{hs}(2,\mathbb{R})$ subalgebra and internal subalgebra}
The polynomials that contain no $\psi_i$ (degree $K$ equal to zero) form a subalgebra, which is nothing but the algebra $\textrm{hs}(2,\mathbb{R})$ that has been used for the description of the integer higher-spin gravity theory \cite{Blencowe:1988gj}.  It is a subalgebra of the bosonic subalgebra containing the polynomials of even $K$-degree.

Another interesting subalgebra is the finite subalgebra of polynomials involving only $\psi$'s and no $q$'s. We call it the internal subalgebra $\textrm{U}$. The internal subalgebra $\textrm{U}$ contains $\textrm{so}(N,\mathbb{R})$ as the subalgebra generated by the quadratic monomials $X_{ij}$.  To identify $\textrm{U}$ completely, we recall that the $\psi_i$'s are the generators of a Clifford algebra. When $N$ is even, the internal subalgebra is therefore the direct sum
\be
\textrm{U} = \textrm{su}(2^{ \frac{N-2}{2}}) \oplus \textrm{su}(2^{ \frac{N-2}{2}})  \oplus \textrm{u}(1) \; \;  \; \; \; \; \; \hbox{($N$ even)},
\ee
while when $N$ is odd, one gets
\be
\textrm{U} = \textrm{su}(2^{ \frac{N-1}{2}}) \; \;  \; \; \; \; \; \hbox{($N$ odd)}.
\ee

\subsubsection{$N=1$ - alternative description}

For $N=1$ supersymmetry, an alternative description of the superalgebra is available. Since there is only one $\psi$, any element of  $\textrm{shs}^\textrm{E}(1|2,\mathbb{R})$ can be decomposed as
\be
f = P_0 + p_1
\ee
where $P_0$ is a Grassman-even polynomial in the $q$'s containing no $\psi$ while $p_1$ is linear in $\psi$ and reads
\be
p_1 = i P_1 \psi.
\ee
Here, $P_1$ is a Grassman-odd polynomial in the $q$'s.  Furthermore, $P_0$ contains only terms of even degrees in the $q$'s while $P_1$ contains only terms of odd degrees.

We can associate to $f$ a polynomial $F$ in the $q$'s with no constant term as follows:
\be
f =P_0 + p_1 \mapsto F = P_0 + P_1.
\ee
Here, $F$ is pure imaginary and contains both even ($P_0$) and odd ($P_1$) powers in the $q$'s.  The even part $P_0$ is also Grassman-even, while the odd part $P_1$ is Grassman-odd.  In terms of this new representation, the $\star$-product reads
\be
\label{starprod}
(F\star G) (q'')\equiv \exp\left(i\,\epsilon_{\alpha\beta}\frac{\partial}{\partial q_\alpha}\frac{\partial}{\partial q'_\beta}\right)F(q)G(q')\,{\vphantom{\frac{1}{2}}\vline}_{\,q = q' = q''}\, ,
\ee
and the $\star$-bracket becomes
\be
\label{com}
\bal
~[F,G]_\star &= 2 i \sin\left(\epsilon_{\alpha\beta}\frac{\partial}{\partial q_\alpha}\frac{\partial}{\partial q'_\beta}\right)(F_0(q)G_0(q') + F_1(q) G_0(q') + F_0(q) G_1(q'))\,{\vphantom{\frac{1}{2}}\vline}_{\,q = q' = q''}  \\
 & \; \; \; \;  + 2 \cos\left(\epsilon_{\alpha\beta}\frac{\partial}{\partial q_\alpha}\frac{\partial}{\partial q'_\beta}\right)(F_1(q) G_1(q'))\,{\vphantom{\frac{1}{2}}\vline}_{\,q = q' = q''} .
\eal
\ee

The $N=1$ super-algebra $\textrm{she}^E(1|2, \mathbb{R})$ is thus isomorphic to the superalgebra $\textrm{shs}(2, \mathbb{R})$, defined to be the superalgebra of  imaginary polynomials in the $q$'s with no constant term but with both even and odd powers (the coefficients of the even - respectively odd - powers being Grassmann-even - respectively Grassmann-odd), equipped with the $\star$-bracket (\ref{com}).  The above basis (\ref{basis0}) becomes in this alternative description 
\be
\label{basis}
X_{(p,q)} \equiv X_{\underbrace{1\dots 1}_p\underbrace{2\dots 2}_q} \equiv \frac{1}{2\,  i \, p! \, q!}(q_1)^p(q_2)^q\,, \qquad p+q \in \mathbb{N}_0 .
\ee
Some useful relations are worked out explicitly in  Appendix \ref{AppB}.

\subsection{Spin}

For any  $N$-extended supersymmetry, the superalgebra $\textrm{shs}^\textrm{E}(N|2,\mathbb{R})$ contains the spacetime algebra $\textrm{sl}(2, \mathbb{R}) \simeq \textrm{su}(1,1)$ under which it decomposes as a direct sum of irreducible representations.  To exhibit this decomposition, it is convenient to write
\be
\textrm{shs}^\textrm{E}(N|2,\mathbb{R}) = \oplus_{j\geq 0} V_{j}
\ee
where $j$ is a non-negative integer or half-integer, and where $V_{j}$ is the vector subspace containing the polynomials in the $q$'s of degree $2j$ (with no restriction on the degree in the $\psi$'s, which are spacetime scalars).  The subspaces $V_j$ are invariant under the action of $\textrm{sl}(2, \mathbb{R})$ and are reducible for $N > 1$.  More precisely,
\begin{eqnarray}
&& V_{0} = D_0 \otimes  {\mathcal E}' \\
&& V_{j} = D_j \otimes  {\mathcal O} \; \; \; (j \hbox{ half-integer } \geq \frac{1}{2} )\\
&& V_j = D_j \otimes {\mathcal E} \; \; \; (j \hbox{  integer } \geq 1 )
\end{eqnarray}
where $D_j$ is the $(2j+1)$-dimensional space of the $\textrm{sl}(2, \mathbb{R})$-spin $j$ irreducible representation.   Furthermore, $ {\mathcal E}$ is the space of polynomials of even degree in $\psi_i$, $ {\mathcal E}'$ is the space of polynomials of even degree in $\psi_i$ with no constant term, and $ {\mathcal O}$ is the space of polynomials of odd degree in $\psi_i$.  The subalgebra $\textrm{sl}(2, \mathbb{R})$ appears in $V_1$, as $D_1$ times the constants.  The subspaces $ {\mathcal E}$, $ {\mathcal E}'$ and $ {\mathcal O}$ have respective dimensions

\begin{center}
\begin{tabular}{|c|c|c|}
  \hline
   & dim  (for $N=0$) &  dim  (for $N>0$)\\
  \hline
  ${\mathcal E}$ & 1 & $2^{N-1}$  \\
  ${\mathcal E'}$ &0 &  $2^{N-1} - 1$ \\
 ${\mathcal O}$ & 0 & $2^{N-1}$ \\
  \hline
\end{tabular}
\end{center}

For $N \leq 1$, the space ${\mathcal E}'$,  and hence also the space $V_0$,  is trivial.  Hence, for $N \leq 1$, the spin-$0$ representation does not occur. Furthermore, the spaces  ${\mathcal O}$ and ${\mathcal E}$ are then one-dimensional, so that the subspaces $V_j$ are irreducible and each value of the spin appearing in the theory is non-degenerate.  Neither of these features holds for $N>1$.

To summarize, one encounters the following higher spin superalgebras as one increases the number $N$ of supersymmetries:
\begin{itemize}
\item $\textrm{shs}^\textrm{E}(0|2,\mathbb{R}) \simeq \textrm{hs}(2, \mathbb{R})$ is the bosonic higher spin algebra involving only integer spins $\geq 1$ (no supersymmetry).  Each value of the spin is non-degenerate.
\item  $\textrm{shs}^\textrm{E}(1|2,\mathbb{R}) \simeq \textrm{shs}(2, \mathbb{R})$ is the higher spin superalgebra for simple supergravity.  It contains osp$(1|2, \mathbb{R}) \supset \textrm{sl}(2, \mathbb{R})$ and has no non-trivial internal subalgebra.  It involves both half-integer and integer spins $\geq \frac{1}{2}$.  Each value of the spin is again non-degenerate.
\item $\textrm{shs}^\textrm{E}(N|2,\mathbb{R})$ is relevant for the extended models.  It involves both half-integer and integer spins $\geq \frac{3}{2}$.  Spin $0$ is degenerate $2^{N-1} -1$ times, while spins $\geq \frac{1}{2}$ are degenerate $2^{N-1}$ times.
\end{itemize}

Later, through the Hamiltonian reduction implemented at infinity by the boundary conditions discussed below, we will show that each $\textrm{sl}(2, \mathbb{R})$-representation $D_j$ yields  a generator of  conformal dimension $j+1$.

\section{Higher-Spin Chern-Simons Super-Gauge Theory}
\label{SuperCS}
\setcounter{equation}{0}

We now turn to the dynamics. The starting point is a doubled Chern-Simons gauge theory, whose super-connection 1-forms are $\Gamma$ taking values in $\textrm{shs}^E(N|2,\mathbb{R})$ and $\overline{\Gamma}$ taking values in $\textrm{shs}^E(M| 2, \mathbb{R})$:
\bea
\Gamma(x; q, \psi) &=& \sum_{m, n, i_1, \cdots, i_N} \rmd x^\mu \Gamma^{m, n; i_1, \cdots, i_N}_\mu  (x) X_{m, n; i_1, \cdots, i_N} \\
\overline{\Gamma}(x; q,\psi) &=& \sum_{m, n, i_1, \cdots, i_M} \rmd x^\mu \overline{\Gamma}^{m, n; i_1, \cdots, i_M}_\mu (x) \overline{X}_{m, n; i_1, \cdots, i_M}.
\eea
They can be decomposed further according to the spinor parity:
\bea
\Gamma^{m, n; i_1, \cdots, i_N}_\mu  (x) &=& \left\{ \begin{array}{cc} A_\mu^{m, n; i_1, \cdots, i_N}(x) & \qquad (m+n = \mbox{even}) \\
\Psi_\mu^{m, n; i_1, \cdots, i_N} (x) & \qquad (m+n = \mbox{odd}) \end{array} \right. , \nonumber \\
\overline{\Gamma}^{m, n; i_1, \cdots, i_M}_\mu (x) &=& \left\{ \begin{array}{cc} \overline{A}_\mu^{m, n; i_1, \cdots, i_M}(x) & \qquad (m+n = \mbox{even}) \\
\overline{\Psi}_\mu^{m, n; i_1, \cdots, i_M} (x) & \qquad (m+n = \mbox{odd}) \end{array} \right. .
\eea
The even parity components are real spacetime Bose fields, while the odd parity components are real spacetime Fermi fields.

The super-gauge transformations of these super-connections are given in terms of a super-gauge 0-form $\Lambda(x; q, \psi)$:
\bea
\delta_\Lambda \Gamma(x; q,\psi)
= \rmd \Lambda(x; q, \psi) +\Gamma(x; q, \psi) \star \Lambda (x; q, \psi) - \Lambda (x; q, \psi)
\star \Gamma (x; q, \psi). \; \; \; \;
\eea
In accordance with the super-connection 1-form, the super-gauge 0-form is expandable as
\bea
\Lambda (x; q, \psi) = \sum_{m, n, i_1, \cdots, i_N} \Lambda^{m, n; i_1, \cdots, i_N}(x)  X_{m, n; i_1, \cdots, i_N}
\eea
where the real coefficients
\bea
\Lambda^{m, n; i_1, \cdots, i_N}(x) = \left\{ \begin{array}{cc} \lambda^{m, n; i_1, \cdots, i_N}(x) & \qquad (m+n = \mbox{even}) \\
\eta^{m, n; i_1, \cdots, i_N}(x) & \qquad (m+n = \mbox{odd}) \end{array} \right.
\eea
parametrize respectively the bosonic and the fermionic gauge transformations.

The theory is defined by the action
\bea
S_{\rm HS} [\Gamma, \overline{\Gamma}] = S_{\rm cs}[\Gamma] - S_{\rm cs}[\overline{\Gamma}].
\eea
with a relative minus sign.  The first part is referred to as the ``chiral sector" whereas the second part is the ``anti-chiral sector".
The Chern-Simons action is given for the chiral part by
\bea
S_{\rm cs}[\Gamma] &\equiv& {k \over 4 \pi} \int_{{\cal M}_3} \mbox{Str} \Big( \Gamma \wedge \rmd \star \Gamma + {2 \over 3} \Gamma \wedge \star \Gamma \wedge \star \Gamma \Big) \nonumber \\
&=& {k \over 4 \pi} \int_{{\cal M}_3}
\Big[ \mbox{Tr} \big(A \wedge \rmd A + {2 \over 3} A \wedge A \wedge A \big) + i \textrm{Tr} \big(\overline{\Psi} \wedge \rmd \Psi + \overline{\Psi} \wedge A \wedge \Psi \big) \Big]
\eea
and similarly for the anti-chiral part.
The coefficient $k$ is a dimensionless, real-valued coupling constant of the theory. In the gravitational context considered here, it is related to the three-dimensional Newton's constant $G$ and the AdS radius of curvature $\ell$ through
\bea
 k = {\ell \over 4 G}. \label{RelationGk}
\eea
The cosmological constant is $\Lambda \equiv - {1 \over \ell^2}$.  With $k$ real, the action is real-valued. 

As discussed in the previous section, the gauge algebra $\textrm{shs}^E(N|2, \mathbb{R}) \oplus \textrm{shs}^E(M| 2, \mathbb{R})$ contains various finite-dimensional subalgebras. When the gauge algebra is restricted to the $\textrm{so}(1, 2, \mathbb{R}) \oplus \textrm{so}(1, 2, \mathbb{R})$ bosonic algebra, the theory is reduced to the Chern-Simons formulation of the three-dimensional Einstein gravity with negative cosmological constant. When the gauge algebra is restricted to the $\textrm{osp}(1| 2, \mathbb{R}) \oplus \textrm{osp}(1| 2, \mathbb{R})$ superalgebra, this theory is reduced to the Chern-Simons formulation of three-dimensional ${\cal N}=(1,1)$ Einstein supergravity with negative cosmological constant.  When the gauge algebra is truncated to $\textrm{sl}(3, \mathbb{R}) \oplus \textrm{sl}(3, \mathbb{R})$ (which is not a subalgebra, but one can proceed along the lines explained in  \cite{Henneaux:2010xg}), the theory is reduced to the Chern-Simons formulation of  three-dimensional spin-3 gravity with negative cosmological constant, which describes the consistent interaction of a spin-3 field with Einstein gravity. In all these cases, the vacuum is the three-dimensional anti-de Sitter space. It is important to note that the isometry algebra $\textrm{sl}(2, \mathbb{R}) \oplus \textrm{sl}(2, \mathbb{R})$ of the vacuum configuration coincides with the gravitational subalgebra  $\textrm{sl}(2, \mathbb{R}) \oplus \textrm{sl}(2, \mathbb{R})$ of the respective gauge algebras.  When Killing spinors are included in the context of (2+1)-dimensional supergravities  \cite{Coussaert:1993jp}, this gravitational algebra is enlarged to the corresponding superalgebras.

Though containing an infinite number of components, the Chern-Simons super-gauge theory has no propagating field degrees of freedom. The field equations
\bea
&& F(\Gamma) \equiv \rmd \Gamma + \Gamma \wedge \star \Gamma = 0 \nonumber \\
&& \overline{F}(\overline{\Gamma}) \equiv \rmd \overline{\Gamma} + \overline{\Gamma} \wedge \star
\overline{\Gamma} = 0
\eea
assert that the super-connections $\Gamma, \overline{\Gamma}$ are flat. This means that locally the connections can be put into a pure-gauge configuration:
\bea
\Gamma (x, \xi) = U^{-1} (x, \xi) \star \rmd U (x, \xi) \qquad \mbox{and} \qquad
\overline{\Gamma} (x, \xi) = \overline{U}^{-1} (x, \xi) \star \rmd \overline{U} (x, \xi).
\eea
The configuration can still leave degrees of freedom describing global charges or holonomies, depending on the geometry and topology of the three-manifold ${\mathcal M}_3$  over which the theory is defined. Unraveling the global charges in the asymptotically AdS background is one main task of this paper.

\section{Asymptotic symmetries}
\label{AsymptoticSymmetries}
\setcounter{equation}{0}

\subsection{Asymptotics  of $\textrm{shs}^\textrm{E}(N|2,\mathbb{R})$ connection}

Consider the spacetime manifold $\mathcal{M}_3$ of topology $\mathbb{R}\times\mathcal{D}$. Here, $\mathbb{R}$ parametrizes the time coordinate and $\mathcal{D}$ is a two-dimensional spatial manifold, which is assumed to have at least one boundary  that we call ``asymptotic infinity'' or more loosely ``infinity''. This boundary is assumed to correspond to $r\rightarrow\infty$, where
spacetime approaches AdS$_3$,
\bea
\rmd s^2 \rightarrow {\ell^2 \over r^2} \left[ - \rmd x^+ \rmd x^- +  \rmd r^2 \right],
\eea
(see Appendix A).

We impose asymptotic conditions on the connection that simultaneously generalize those of \cite{Henneaux:2010xg} for higher spin bosonic models and those of \cite{Banados:1998pi,Henneaux:1999ib} for simple and $N$-extended supergravities.

In the case of minimal AdS$_3$ supergravity, the boundary conditions  were  \cite{Banados:1998pi} (after an appropriate gauge transformation that simplifies the form of the connection and its  $r$-dependence \cite{Coussaert:1995zp})
\bea
&& \Gamma (x) \rightarrow [- 1 \cdot X_{22} + B^1_+ (x^+, x^-) X_1 + B^{11}_+ (x^+, x^-) X_{11} ] \rmd x^+
 \\
&& \overline{\Gamma}(x) \rightarrow [+ 1 \cdot X_{11} + \overline{B}^2 (x^+, x^-) X_2 + \overline{B}^{22}_- (x^+, x^-) X_{22}] \rmd x^- .
\eea
In the case of higher-spin AdS$_3$ gauge theory, the boundary conditions were \cite{Henneaux:2010xg}
\bea
&& \Gamma (x) \rightarrow [- 1 \cdot X_{22} + \Delta^{11} (x^+, x^-) X_{11} + \Delta^{1111} (x^+, x^-) X_{1111} + \cdots] \rmd x^+ \\
&& \overline{\Gamma} (x) \rightarrow [+ 1 \cdot X_{11} + \overline{\Delta}^{22}(x^+, x^-) X_{22}
+ \overline{\Delta}^{2222} (x^+, x^-) X_{2222} + \cdots] \rmd x^-.
\eea
Combining these two limiting situations, it is fairly obvious that the correct boundary conditions for the $\textrm{shs}(1| 2, \mathbb{R})$-valued gauge connections of simple supergravity are
\bea
&& \Gamma(x) \rightarrow \big[ - 1 \cdot X_{22} + \sum_{\ell = 1}^\infty \Delta^{(\ell, 0)} (x^+, x^-) X_{(\ell, 0)} \big] \rmd x^+ \\
&& \overline{\Gamma}(x) \rightarrow \big[ + 1 \cdot X_{11} + \sum_{\ell = 1}^\infty \overline{\Delta}^{(0, \ell)} (x^+, x^-) X_{(0, \ell)} \big] \rmd x^- .
\eea

The boundary conditions for the theories with extended supersymmetry are similar but without imposing a highest-weight or lowest-weight type of gauge condition along the internal symmetry algebra.   Indeed, it was found in \cite{Henneaux:1999ib} that the requisite boundary conditions for extended supergravities took the form
\bea
&& \Gamma (x) \rightarrow [- 1 \cdot X_{22} + B^{1i}_+ (x^+, x^-) X_{1i} + B^{11}_+ (x^+, x^-) X_{11} + B^{ij}_+ (x^+, x^-) X_{ij}] \rmd x^+
\\
&& \overline{\Gamma}(x) \rightarrow [+ 1 \cdot X_{11} + \overline{B}^{2i}_- (x^+, x^-) X_{2i} + \overline{B}^{22}_- (x^+, x^-) X_{22} + \overline{B}^{ij}_- (x^+, x^-) X_{ij}] \rmd x^-
\eea
with no restriction on the internal indices occurring asymptotically.  Therefore, we impose
\bea
&& \Gamma (x) \rightarrow [- 1 \cdot X_{22} + \sum \Delta^{pi_1 \cdots i_N} (x^+, x^-) X_{p,0;i_1 \cdots i_N}] \rmd x^+  \label{AsCo+}
\\
&& \overline{\Gamma}(x) \rightarrow [+ 1 \cdot X_{11} + \sum \overline{\Delta}^{qi_1 \cdots i_N} (x^+, x^-) X_{0,q;i_1 \cdots i_N} ] \rmd x^-  \label{AsCo-}
\eea
where we sum on repeated indices over all their possible values. Note in particular that the values $p=0$ and $q=0$  occur when the degree  $K = i_1 + i_2 + \cdots + i_N$ does not vanish.

Even though there is no asymptotic restriction on the weights of the representations of the internal algebra, we continue to call the boundary conditions (\ref{AsCo+}) and (\ref{AsCo-}) the ``highest-weight", respectively, the ``lowest-weight" gauge boundary conditions, in analogy with the non-extended cases ($N= 0$ or $N=1$).

\subsection{Hamiltonian reduction}

The above boundary conditions on the super-connections coincide with the constraints that implement the familiar Drinfeld-Sokolov (DS) Hamiltonian reduction \cite{Drinfeld:1984qv,Bouwknegt:1992wg} in the WZWN models  \cite{Alekseev:1988ce,Bershadsky:1989mf,Bershadsky:1989tc,Forgacs:1989ac,Balog:1990mu,Bais:1990bs} -- to which the Chern-Simons theory reduces on the boundary \cite{Elitzur:1989nr}.  As it has been demonstrated in those references, the Virasoro algebra (or one of its appropriate extensions) emerges in the reduction procedure from the current algebra of the unreduced theory.

That the AdS$_3$ boundary conditions implement the DS Hamiltonian reduction was pointed out first in the case of pure AdS$_3$ gravity in \cite{Coussaert:1995zp}, where the Virasoro algebra  is generated from the affine $\textrm{sl}(2, \mathbb{R})$ current algebra (one in each chiral sector).  This was then extended to the case of $N=1$ supergravity, where one gets after reduction the $N=1$ superconformal algebra \cite{Banados:1998pi}, and further to extended supergravity theories in \cite{Henneaux:1999ib}.  In that latter case, the extended superconformal algebras that arise contain nonlinearities in the Kac-Moody currents, realizing the algebraic structures uncovered in \cite{Knizhnik:1986wc,Bershadsky:1986ms,Fradkin:1991gj,Fradkin:1992bz,Fradkin:1992km,Bowcock:1992bm,Bina:1997bm}.

In all these cases, the conformal dimensions of the generators of the boundary superconformal algebras are $\leq 2$  because the underlying bosonic algebras in the bulk are of the form $\textrm{sl}(2,\mathbb{R}) \oplus \mathcal{G}$ and the $\textrm{sl}(2,{\mathbb R})$-representations involve only spins $\leq 1$.  The analysis was more recently generalized to include higher conformal dimensions in \cite{Henneaux:2010xg} and \cite{Campoleoni:2011hg} with bulk algebras being the infinite-dimensional $\textrm{hs}(2, \mathbb{R})$ and the finite-dimensional $\textrm{sl}(N, \mathbb{R})$, respectively.

Because the boundary conditions (\ref{AsCo+}) and (\ref{AsCo-}) are precisely those that implement the Hamiltonian reduction of affine superalgebras, one can proceed along the well-known DS reduction  \cite{Bouwknegt:1992wg} to derive the corresponding asymptotic symmetry algebras.  The precise steps adapted to an infinite number of AdS$_3$ spins have been given in  \cite{Henneaux:2010xg}.  We shall follow this reference here, stressing the conceptual points rather than giving explicit formulas, which are rather cumbersome indeed.  However, the machinery to derive systematically the formulas will be explained.

\subsection{Residual gauge transformations}

Given theAdS$_3$ boundary conditions (\ref{AsCo+}) and (\ref{AsCo-}), the next step is to look for the residual gauge transformations that act nontrivially at asymptotic infinity while leaving the boundary conditions intact. With gauge parameter $\Lambda(x)$, the infinitesimal gauge transformation of $\Gamma$ reads
\bea
\Gamma \rightarrow \Gamma' = \Gamma + \delta \Gamma, \qquad \mbox{where} \qquad
\delta \Gamma = d \Lambda + [\Gamma, \Lambda].
\eea
We see that, in order for $\Gamma'$ to retain the given asymptotics, $\Lambda$ cannot possibly depend on $r$ or $x^-$ to leading order at infinity. Moreover, the gauge transformations should not generate any other components than the highest-weight ones already present. A similar argument goes for $\overline{\Gamma}$. With gauge parameter $\overline{\Lambda}(x)$, the infinitesimal gauge transformation of $\overline{\Gamma}$ reads
\bea
\overline{\Gamma} \rightarrow \overline{\Gamma}' = \overline{\Gamma} + \delta \overline{\Gamma},
\qquad \mbox{where} \qquad
\delta \overline{\Gamma} = d \overline{\Lambda} + [\overline{\Gamma}, \overline{\Lambda}].
\eea
Again, in order for $\overline{\Gamma}'$ to retain the boundary condition (\ref{AsCo-}), $\overline{\Lambda}$ cannot possibly depend on $r$ or $x^+$. Furthermore, the gauge transformations should not generate any other components than the lowest-weight ones already present in  (\ref{AsCo-}).
Summarizing, we found that the gauge transformations $\Lambda(x^+)$ and $\overline{\Lambda}(x^-)$ must be chiral, respectively, antichiral at the least. These functions must be subject to further conditions in order to retain the boundary conditions. This is the task we will undertake next, treating explicitly for definiteness the positive chirality sector (the negative chirality sector is treated similarly).

To proceed further, we find it convenient to decompose the gauge transformations in stacks of successively higher $\textrm{sl}(2, {\mathbb R})$-spin layers. This is because, for each spin, the highest-weight or the lowest-weight components are the only ones that appear in the boundary conditions for the gauge connection. We thus write
\bea
\Lambda (x^+) &=&
\sum_{m, n, i_1, \cdots, i_N}  \Lambda^{m, n; i_1, \cdots, i_N}  (x^+) X_{m, n; i_1, \cdots, i_N}
 \nonumber \\
&=& \Lambda^{\textrm{LW}} + \lambda
\eea
with
\bea
\Lambda^{\textrm{LW} } &= & \sum_{i_1 + \cdots + i_N \geq 2} \Lambda^{0, 0; i_1, \cdots, i_N}(x^+) X_{0, 0; i_1, \cdots, i_N} + \sum_{i_1 + \cdots + i_N \geq 1} \Lambda^{0, 1; i_1, \cdots, i_N}(x^+) X_{0, 1; i_1, \cdots, i_N} \nonumber \\
&&+ \sum_{\ell = 2}^\infty \sum_{i_1, \cdots, i_N}\Lambda^{0, \ell; i_1, \cdots, i_N}(x^+) X_{0, \ell; i_1, \cdots, i_N}
\eea
and
\bea
\lambda &=&   \sum_{i_1 + \cdots +i_N \geq 1} \Lambda^{1, 0; i_1, \cdots, i_N}(x^+) X_{1, 0; i_1, \cdots, i_N}  + \sum_{\ell = 2}^\infty \sum_{i_1, \cdots, i_N}\Lambda^{1, \ell-1; i_1, \cdots, i_N}(x^+) X_{1, \ell -1; i_1, \cdots, i_N} \nonumber \\
&& +  \sum_{\ell = 2}^\infty \sum_{i_1, \cdots, i_N}\Lambda^{2, \ell-2; i_1, \cdots, i_N}(x^+) X_{2, \ell -2; i_1, \cdots, i_N} +  \cdots \nonumber \\
&&+ \cdots + \sum_{\ell \geq s }^\infty \sum_{i_1, \cdots, i_N}\Lambda^{s, \ell-s; i_1, \cdots, i_N}(x^+) X_{s, \ell -s; i_1, \cdots, i_N}  + \cdots.
\eea
In plain words, we collected all the lowest-weight states, which are the states involving  $X_{0, s; i_1, \cdots, i_N}$ in $\Lambda^{LW}$ and, at the same time,  all higher weight states, involving $X_{m, n; i_1, \cdots, i_N}$ with $m>0$, are packaged together in $\lambda$. We should also stress that, although this is not written explicitly, the sums in the above expressions are always restricted to total even degree.  So, for instance, $i_1 + \cdots +i_N$ must be even in the first term in the right-hand side of the expression for $\Lambda^{\textrm{LW} }$, while it must be odd in the second term.  Such a convention will always be adopted in the sequel.

The reason for proceeding in this manner is that the requirement that the asymptotic boundary conditions be preserved determines $\lambda$ in terms of $\Lambda^{LW}$.   Indeed, let us compute $\delta \Gamma = \rmd \Lambda + [\Gamma, \Lambda]$.
Structurally,
\bea
\delta \Gamma = {\sum_{m, n; i_1, \cdots, i_N}} \gamma^{m, n; i_1, \cdots, i_N}(x^+) X_{m, n; i_1, \cdots, i_N}
\eea
where
\bea
\gamma^{m, n; i_1, \cdots, i_N}(x^+) = \partial_+ \Lambda^{m, n; i_1, \cdots, i_N} + [\Gamma, \Lambda]^{m, n; i_1, \cdots, i_N}.
\eea
Since the only non-vanishing components of $\Gamma$ at infinity are $\gamma^{m, 0; i_1, \cdots, i_N}$ (apart from $\gamma^{0, 2; 0, \cdots, 0}$, which is fixed to be equal to $-1$),  the requirement that these global gauge transformations do not alter the boundary conditions is that
\bea
\gamma^{m, 1; i_1, \cdots, i_N} = \gamma^{m, 2; i_1, \cdots, i_N} = \cdots = 0 \qquad \mbox{for} \qquad m =0, 1, 2, \cdots
\label{stackcond1}
\eea
or, equivalently,
\bea
\gamma^{s, \ell - s; i_1, \cdots, i_N} = 0 \qquad \mbox{for} \qquad \ell \ge s+1, \quad s = 0, 1, 2, \cdots.
\label{stackcond2}
\eea
The highest-weight terms $\gamma^{m, 0; i_1, \cdots, i_N}$ are not constrained to be zero and are equal to $\Delta^{mi_1 \cdots i_N}$ according to  (\ref{AsCo+}).

Now, since
$$
[X_{22}, X_{m, n; i_1, \cdots, i_N} ] \sim X_{m-1, n+1; i_1, \cdots, i_N} \; \; \; (m\geq 1)
$$
one may solve recursively the conditions for the higher-weight coefficients $\Lambda^{1, n; i_1, \cdots, i_N}$, $\Lambda^{2, n; i_1, \cdots, i_N}$, ..., given the lowest-weight ones $\Lambda^{0, k; i_1, \cdots, i_N}$, along exactly the same lines as developed in \cite{Henneaux:2010xg}.  One starts from the lowest-weight conditions $\gamma^{0, \ell; i_1, \cdots, i_N} = 0$ ($\ell \geq 1$) to determine the level-one coefficients $\Lambda^{1, \ell-1; i_1, \cdots, i_N}$.  Then one proceeds to solving the level-one conditions  $\gamma^{1, \ell-1; i_1, \cdots, i_N} = 0$ ($\ell \geq 2)$) to determine the level-two coefficients $\Lambda^{2, \ell-2; i_1, \cdots, i_N}$.  One walks one's way up step by step in this fashion. The last set of conditions $\gamma^{\ell-1, 1; i_1, \cdots, i_N} = 0$  ($\ell \geq 1$) determine the highest-weight coefficients $\Lambda^{\ell, 0; i_1, \cdots, i_N}$.  

It should be stressed that, during the process, the higher-weight coefficients depend not only on the lowest-weight coefficients but also on their derivatives.  To emphasize this feature, we shall say that the higher-weight coefficients are {\it functionals} of the lowest-weight ones.  The solutions depend also on the (non-zero) coefficients of the connection and their derivatives.

Collecting the results of the above structure analysis, we conclude that the gauge transformations that leave the
boundary conditions intact are completely specified by the lowest-weight components $\Lambda^{0, k; i_1, \cdots, i_N}$ of the gauge function, while all higher-weight components are determined functionally in terms of these lowest-weight components of the gauge function and the highest-weight components of the original gauge connection.
Notice that, as in the higher-spin bosonic case as well as in the extended supergravity models, the solution for the higher-weight components of the gauge function $\Lambda$ in terms of the lowest-weight ones,  the free gauge potential components $\Delta^{mi_1 \cdots i_N}$ and their derivatives is nonlinear.  It is this feature that will render the resulting asymptotic algebra also nonlinear.

\subsection{Asymptotic symmetry superalgebra}
To identify the asymptotic symmetry superalgebra, one needs to extract the commutation relations for the superalgebra of asymptotic gauge transformations induced by the gauge function $\Lambda$. In the canonical formalism, these commutation relations are realized as the Poisson brackets of the generators of these asymptotic symmetries (up to possible central charges \cite{Brown:1986ed}), and we shall focus on these here.

Consider a phase-space observable ${\cal O}$. Under the global gauge transformation parametrized by $\Lambda$, this observable transforms according to
\bea
{\cal O} \rightarrow {\cal O} + \delta {\cal O} \quad \mbox{with} \quad
\delta {\cal O} = \{ {\cal O}, G[\Lambda] \}_{\rm PB}.
\label{generalgaugetransf}
\eea
On an equal-time slice $\Sigma_2$, the functional of gauge transformation $G[\Lambda]$ is given by
\bea
G[\Lambda] = \int_{\Sigma_2} \sum_{m, n, i_1, \cdots, i_N}  \Lambda^{m, n; i_1, \cdots, i_N} {\mathcal G}_{m, n; i_1, \cdots, i_N} + S_\infty,
\eea
where ${\mathcal G}_{m, n; i_1, \cdots, i_N}$ are the Gauss law constraints and $S_\infty$ is a boundary term at asymptotic infinity defined by the requirement that $G[\Lambda]$ must have well-defined functional derivatives with respect to the connection components, i.e., $G[\Lambda]$ must be such that
$\delta G[\Lambda]$ contains only undifferentiated field variations under the given boundary conditions for $\Gamma$ \cite{Regge:1974zd}.

In the present case, the on-shell configuration is
\bea
{\cal G} = 0,
\eea
so that the generator reduces on-shell to the surface term $S_\infty$.  On the other hand, $S_\infty$ just follows from straightforward integration by part and is proportional to the angular components of the connection along the highest weight basis vectors times the components of the gauge parameters along the lowest weight basis vectors (to leading order), as it was already found for supergravity \cite{Banados:1998pi,Henneaux:1999ib}.  Explicitly,
\be
S_\infty 
=  \oint  \sum_{s, i_1, \cdots, i_N}  \Lambda^{0, s; i_1, \cdots, i_N} \Delta^{s}_{\; \; \;  i_1, \cdots, i_N} \, ,
\ee
where we have redefined the $\Delta$'s through the absorption of the factors  that appear in front of the integral, which we denote by $\alpha_{s, 0; i_1, \cdots, i_N}$, $$\Delta^{ s; i_1, \cdots, i_N} = \Gamma^{s, 0; i_1, \cdots, i_N} \alpha_{s, 0; i_1, \cdots, i_N}. $$  We thus see that (up to those factors) the generators of the asymptotic symmetries are indeed nothing but the leading terms  in the asymptotic expansion of the highest-weight components $\Gamma^{s, 0; i_1, \cdots, i_N}$ of the gauge connection.

The algebra of the asymptotic symmetry generators $\Delta^{ s; i_1, \cdots, i_N} $can be read off by equating their variations under an arbitrary asymptotic symmetry transformation, computed in two different ways.  First, $\delta  \Delta^{ s; i_1, \cdots, i_N}$ can be derived from the gauge variation formula,
\be
\delta\Delta^{ s; i_1, \cdots, i_N}(\theta) =  \delta \Gamma^{s, 0; i_1, \cdots, i_N}(\theta)\,\alpha_{s, 0; i_1, \cdots, i_N} = \left(\partial\Lambda^{ s,0; i_1, \cdots, i_N} + [\Gamma, \Lambda]^{ s,0; i_1, \cdots, i_N} \right) \alpha_{s, 0; i_1, \cdots, i_N}\ ,
\ee
with the $\Lambda^{ m,n; i_1, \cdots, i_N}$ determined from the lowest-weight $\Lambda^{ 0,s; i_1, \cdots, i_N}$ along the lines explained in the previous subsection.  Second, $\delta  \Delta^{ s; i_1, \cdots, i_N}$ can be obtained directly from (\ref{generalgaugetransf}),
\be
\delta \Delta^{ s; i_1, \cdots, i_N}(\theta) = \{ \Delta^{ s; i_1, \cdots, i_N}(\theta), \oint  \sum_{s', j_1, \cdots, j_N}  \Lambda^{0, s'; j_1, \cdots, j_N} \Delta^{ s'}_{\; \; \; j_1, \cdots, j_N}(\theta)\}_{\rm PB} \, .
\ee
Here, $\theta$ denotes the angular coordinate of the asymptotic infinity. Equating these two ways of computing  $ \delta\Delta^{ s; i_1, \cdots, i_N}$ yields the Poisson brackets
\bea
\{ \Delta^{ s; i_1, \cdots, i_N}(\theta), \Delta^{ s'; j_1, \cdots, j_N}(\theta') \}_{\rm PB} \qquad
\mbox{for} \qquad s, s' \in \mathbb{N} \, .
\eea

It is evident that this algebra is closed, since the variations $\delta \Gamma^{s, 0; i_1, \cdots, i_N} = \gamma^{s, 0; i_1, \cdots, i_N}$, determined through the recursive procedure explained above,  are functionals of $\Gamma^{s,0; i_1, \cdots, i_N} \sim \Delta^{s; i_1, \cdots, i_N}$ only (in addition to depending linearly on the independent gauge parameters $\Lambda^{0,s; i_1, \cdots, i_N} $).   The functional dependence of $\gamma^{s, 0; i_1, \cdots, i_N}$ on $\Delta^{s; i_1, \cdots, i_N}$ is nonlinear, which implies that the algebra of the $\Delta$'s is nonlinear.  The terms independent of $\Delta$ and linear in the gauge parameters corresponds to the central charges.
Although nonlinear, the algebra obeys of course the Jacobi identity since the Poisson bracket does\footnote{Upon gauge fixing, the Poisson algebra becomes the Dirac algebra. However, the asymptotic algebra does not depend on the gauge choice because the constraints of the theory are all first class.}.  

\section{Nonlinear Super-$W_\infty$ Algebra}
\label{NonlinearW}
\setcounter{equation}{0}

The actual computation of the algebra $\mathcal{SW}$ of the $\Delta^{s; i_1, \cdots, i_N}$'s is rather cumbersome but it can be identified to be a super-$W_\infty$ by following a general argument similar to the one given in \cite{Henneaux:2010xg} for the bosonic case.  We consider first the $N=1$ case, i.e., $\textrm{shs}^E(1|2, \mathbb{R})$:
\begin{enumerate}
\item By computing the general solution to the equations for the $\Lambda^{(m,n)}$'s ($m > 0$) when only the free gauge parameter $\Lambda^{(0,2)}$ is non zero, one observes (i) that the generators $L \equiv \Delta^2$  form a Virasoro algebra with central charge $k/4 \pi$:
\begin{equation}
\{L(\theta), L(\theta')\}_{\rm PB} = \frac{k}{4 \pi}  \delta'''(\theta - \theta') - \left(L(\theta) + L(\theta') \right)  \delta'(\theta - \theta'); \label{LLGen}
\end{equation}
and (ii) that the generators $M_{\frac{j}{2}+1} \equiv \Delta^j$ have conformal dimension $(\frac{j}{2}+1)$:
\begin{equation}
\{L(\theta), M_{\frac{j}{2}+1}(\theta')\}_{\rm PB}  =  -\left( M_{\frac{j}{2}+1}(\theta) + \frac{j}{2} M_{\frac{j}{2}+1}(\theta') \right)  \delta'(\theta - \theta') \ . \label{LMGen}
\end{equation}
\item By finding the general solution to the equations for the $\Lambda^{(m,n)}$'s ($m > 0$) when only the free (fermionic) gauge parameter $\Lambda^{(0,1)}$ is non-zero, one observes that the generator $Q \equiv M_{\frac{3}{2}} \equiv \Delta^1$  is the supercharge,
\bea
i \{ Q(\theta), M_{\frac{s}{2} +1} (\theta') \}_{\rm PB}
&=& - {k \over  \pi} \delta_{s,1} \delta''(\theta - \theta') \nonumber \\
+&& (s + 1) M_{\frac{s+3}{2}} (\theta) \; \; \; \hbox{ ($s$ odd)}
\eea
and
\be
 \{ Q(\theta), M_{\frac{s}{2} +1}(\theta') \}_{\rm PB}
=
  - \delta'(\theta - \theta') \left({1 \over s} M_{\frac{s+1}{2}}(\theta) + M_{\frac{s+1}{2}}(\theta') \right) \; \; \; \hbox{ ($s$ even)}.
\ee
\end{enumerate}

We see that the relations are linear at these levels.  They start to display the nonlinear structure of the algebra at higher levels. For instance, one finds explicitly %
\bea
\{ M_{\frac{5}{2}} (\theta), M_{\frac{5}{2}} (\theta') \}_{\rm PB} &=& {\alpha^3 \over 6} \delta''''(\theta - \theta') + {\alpha^3 \over 12\alpha^6}
(N^6(\theta) + N^6(\theta')) \delta(\theta - \theta') \nonumber \\
&+& {3\alpha^3 \over 2(\alpha^2)^2} L(\theta)L(\theta') \delta (\theta - \theta') - {5\alpha^3 \over 6\alpha^2} \delta'' (\theta - \theta
')(L(\theta) + L(\theta')) \nonumber \\
&-& {\alpha^3 \over 3\alpha^2} \delta'(\theta - \theta')(L'(\theta) - L'(\theta')) + {i\alpha^3 \over 6(\alpha^1)^2} Q(\theta) Q(\theta') \delta' (\theta - \theta') \ \ \ \ \
\eea
and
\bea
\{M_3(\theta), M_3(\theta') \}_{\rm PB}
&=&  {\alpha^3 \over 24} \delta'''''(\theta - \theta') - {5\alpha^3 \over 12\alpha^2} (L(\theta) +L(\theta')) \delta'''(\theta - \theta') \nonumber \\
&+& {\alpha^3 \over 6\alpha^6} (N^6(\theta) + N^6(\theta') ) \delta'(\theta - \theta') + {2i\alpha^3 \over 3(\alpha^1)^2} Q(\theta)Q(\theta') \delta''(\theta - \theta') \nonumber \\
&-& \frac{i\alpha^3}{2(\alpha^1)^2} (Q'(\theta)Q(\theta) + Q'(\theta')Q(\theta')) \delta'(\theta - \theta') \nonumber \\
&+& {\alpha^3 \over (\alpha^2)^2} \delta'(\theta - \theta') (L^2(\theta) + \frac{2L(\theta)L(\theta')}{3}  + L^2(\theta')) \nonumber \\
&+& \frac{\alpha^3}{4\alpha^2}\delta''(\theta-\theta')(L'(\theta') - L'(\theta)) .
\eea

The numerical factors $\alpha^i$ appearing in these expressions read
\begin{equation}
\alpha^i = \frac{k(-)^{n+1}i^n}{\pi n!}.
\end{equation}

For  $N \ge 2$ extended supersymmetry, the derivation proceeds essentially in the same way.  The salient new features that arise are:
\begin{enumerate}
\item There are now fields $\Delta^{0; i_1, \cdots, i_N}$ of conformal dimension $1$. These are the currents of the internal symmetry, and they form an affine subalgebra.   Their brackets with the other generators reflect how these other generators transform under the internal symmetry.  Indeed,  when the only non-vanishing lowest-weight free components are $\Lambda^{0,0; i_1, i_2, \cdots, i_N}$ ($i_1 + \cdots i_N \geq 2$),  the solution for $\Lambda$ is found to be simply $\Lambda = \Lambda^{\textrm{LW} } =  \sum_{i_1 + \cdots i_N \geq 2} \Lambda^{0, 0; i_1, \cdots, i_N} X_{0, 0; i_1, \cdots, i_N} $.
\item A Sugawara redefinition of the Virasoro generator $L$ must actually be performed, as already found in \cite{Henneaux:1999ib} for the extended AdS$_3$ supergravity (see that reference for details).
\item While there is a single generator $M_j$ at each conformal dimension $>1$ for  $N=1$, this is not any more the case for $N \ge 2$.  The degeneracy of each conformal dimension $>1$ is equal to $2^{N-1}$, while the degeneracy of conformal dimension $1$ is $2^{N-1} -1$.  In particular, the Virasoro generator is not the only field with conformal dimension $2$ for extended supersymmetries.
\end{enumerate}

We stress that our construction guarantees automatically that the brackets among the generators fulfill the Jacobi identity since these are just Poisson brackets (or Dirac brackets  if one fixes the gauge). This is worth emphazising since other methods for constructing super $W$-algebras met with difficulties with the Jacobi identity.

Although there is no consistent truncation of  $\textrm{shs}(N|2, \mathbb{R})$ to finite dimensional superalgebras that can be made beyond  $\textrm{osp}(N\vert 2, \mathbb{R})$, the Hamiltonian reduction procedure is very similar to that encountered for the finite-dimensional superalgebra $\textrm{sl}(n+1 | n)$, which yields $N=2$ supersymmetric models with generators $M_s$ of higher conformal dimensions up to 
$s = \frac{2n+1}{2}$ \cite{FigueroaO'Farrill:1991bd,Komata:1990cb,Inami:1990hk,Inami:1991af}.

\section{shs$^E(N| 2, \re)$ as wedge subalgebra of super-W$_\infty$}
\label{wedge}
\subsection{Exact symmetries of the AdS$_3$ background}
In the previous section, we showed that the asymptotic symmetry algebra provides an enhancement from the naive global gauge symmetry algebra shs$^E(N| 2, \re)$ to the super-W$_\infty$ algebra. To deepen our understanding of this remarkable feature, we would like to identify the way in which
the shs$^E(N\vert 2, \re)$ algebra is embedded in the super-W$_\infty$ algebra. In this section, we carry out this analysis in detail.

\subsubsection{Exact symmetries of the zero connection}

The exact symmetry algebra of the AdS$_3$ background is $\textrm{shs}^E(N|2, \re)$.  Indeed, the AdS$_3$ connection is gauge equivalent to zero (it is pure gauge). The zero connection is clearly invariant under gauge transformations that are constant but otherwise arbitrary:
\bea
 O \rightarrow S^{-1}\star \rmd S + S^{-1} \star 0 \star S = 0  \; \; \; \;  \hbox{iff } \; \rmd S =0 \quad \hbox{ viz. } S = S_0,
\eea
with $S_0$ a constant function.  The algebra of constant gauge transformations $S_0$ is of course isomorphic to $\textrm{shs}^E(N,2\vert \re)$.

\subsubsection{The anti-de Sitter connection}
If we denote by $\Gamma^{\textrm{AdS}}$ the AdS$_3$ connection in the standard static-polar reference frame,  we can express it as
\begin{equation}
\Gamma^{\textrm{AdS}} = U^{-1} \, \rmd U .  \label{AdS}
\end{equation}
Here, $U$ is given by the simple expression
\be
U = \exp \left(-\frac{x^+}{2}( X_{11} + X_{22})\right)
\ee
which contains only generators of the $\textrm{sl}(2, \re)$-subalgebra (and neither higher-spin generators nor generators with oscillators' color index).  The generator $X_{11} + X_{22}$ is the compact generator $E-F$ in the Chevalley basis (see appendix B) and generates $\textrm{SO}(2)$.

A few comments are in order:
\begin{itemize}
\item $U$ involves  also $\exp \left( f(r) X_{12} \right)$ for some definite function $f(r)$ [20] but this gauge transformation is irrelevant for the present considerations, so we drop it.  Then, the AdS$_3$ connection  reads
\be 
\Gamma^{\textrm{AdS}} = - \frac{1}{2} \left(X_{11} + X_{22} \right) dx^+  \ . \label{AdS00}
\ee
\item We shall focus on an equal time slice, which we can assume to be $x^0 = 0$, and so we set $x^+ = \varphi$.
\item The transformation $U$ is in the $\textrm{SL}(2, \re)$-subgroup generated by the $X_{\alpha \beta}$ (even in its $\textrm{SO}(2)$-subgroup) and so is the direct sum of the $2 \times 2$ matrix $R$, 
\be
R = \left( \begin{array}{cc}
\cos \frac{\varphi}{2} & -\sin \frac{\varphi}{2}  \\
\sin \frac{\varphi}{2} &\  \cos \frac{\varphi}{2}   \end{array} \right)
\ee
with trivial identity terms in the complementary subspaces.  Note that  $R^{-1}$ is given by
\be
R^{-1} = \left( \begin{array}{cc}
\cos \frac{\varphi}{2} & \sin \frac{\varphi}{2}  \\
-\sin \frac{\varphi}{2} & \ \cos \frac{\varphi}{2}   \end{array} \right)
\ee
and
\be R^{-1} dR = \left( \begin{array}{cc}
0 & - \frac{1}{2}  \\
 \frac{1}{2} & \ \ 0   \end{array} \right)  \ .
\ee
\item To match (\ref{AdS00}) with the asymptotic behavior we were taking, a further  constant gauge transformation $T$ must actually be performed with
\be
T = \exp \left( - \sqrt{2} X_{12} \right) = \left( \begin{array}{cc}
\sqrt{2} &0  \\
0 &\frac{1}{\sqrt{2}}   \end{array} \right).
\ee
It then follows
\be
(RT)^{-1} \rmd (RT) = \left( \begin{array}{cc}
0 &-\frac{1}{4} \\
1 &\ \ 0   \end{array} \right) = - X_{22} - \frac{1}{4} X_{11}
\ee
and this fulfills the asymptotic condition (4.2). The group element $T$ can be combined with $\exp \left( f(r) X_{12} \right)$ above.  The motivation for including $T$ is not only that it makes the coefficient of $-X_{22} \equiv F$ equal to one, but that the connection corresponding to the zero mass black hole  is then simply given by 
$$
\left( \begin{array}{cc}
0 \ & \ 0 \\
1 \ & \ 0   \end{array} \right) = - X_{22} 
$$
in that gauge.   However, for the analysis of this section, we find it more convenient not to include $T$ so that the group element is in $\textrm{SO}(2)$.  The effect of $T$ is to rescale $q_1$ by $\sqrt{2}$ and $q_2$ by $\frac{1}{\sqrt{2}}$, a transformation that does not mix components of different $\textrm{sl}(2,\re)$-weights.  For that reason, the asymptotic analysis we made in the previous section remains unchanged if we do not include $T$.
\end{itemize}

\subsubsection{Exact symmetries of the anti-de Sitter connection}

We can rewrite the constant gauge transformations that leave the zero connection invariant in the representation  where the connection takes the form (\ref{AdS}).  These gauge transformations  are just
\be
S = U^{-1} S_0 \ U \ ,
\ee
where $S_0$ is constant, $\rmd S_0 = 0$.  
In infinitesimal form, $S = I + \Lambda^{\textrm{AdS}}$ with
\be
\Lambda^{\textrm{AdS}} = \sum_{m, n, i_1, \cdots, i_N} \Lambda^{m, n; i_1, \cdots, i_N}_0  U^{-1} X_{m, n; i_1, \cdots, i_N} U \label{LAdS} \ ,
\ee
where $\Lambda^{m, n; i_1, \cdots, i_N}_0$ are constants. For these gauge transformations, 
\be \delta \Gamma^{\textrm{AdS}} = \rmd \Lambda^{\textrm{AdS}} + [ \Gamma^{\textrm{AdS}}, \Lambda^{\textrm{AdS}}] = 0
\ee
It is again obvious that the algebra $[\Lambda^{\textrm{AdS}}_1, \Lambda^{\textrm{AdS}}_2] $ of the exact symmetries of the AdS$_3$ super-connection is $\textrm{shs}^E(N,2\vert \re)$.

\subsubsection{Explicit analysis in terms of lowest-weight components}
The algebra elements $U^{-1} X_{m, n; i_1, \cdots, i_N} U$ can be expanded in the basis of $X_{m, n; i_1, \cdots, i_N} $.  In particular, since we have observed that the lowest $\textrm{sl}(2,\re)$-weight components of the gauge transformations play a central role, we find it interesting to work out the components of  (\ref{LAdS}) along the lowest-weight generators $X_{0, \ell; i_1, \cdots, i_N}$.

To that end, we observe that, as $U$ belongs to the $\textrm{SO}(2)$-subalgebra of $\textrm{SL}(2,\re)$, it does not mix different spins and just acts on the generators $X_{m, n; i_1, \cdots, i_N}$, $m+n= \ell = 2s$ of the spin $s$ representation by the symmetrized $\ell$-th tensor power of the rotation matrix $R$, without affecting the internal indices $i_k$.  The formulas are more transparent if we drop the passive (and hence irrelevant for the present considerations) indices $i_k$ and work in the basis $Y_{m, n}$ with
\be 
Y_{m, n} \equiv z^{m} \bar{z}^n, \quad \mbox{where} \quad z = q_1 + i q_2, \; \; \; \;  \bar{z} = q_1 - i q_2.
\ee
Let us name $\Xi^{m,n}_0$ the coefficients in the new basis.  For the spin $s$ ($= \frac{\ell}{2}$) subspace,
\be
\sum_{m+n= \ell} \Lambda^{m,n}_0 X_{m,n} = \sum_{m+n= \ell} \Xi^{m,n}_0 Y_{m,n} \ . \label{CoB}
\ee
Therefore, 
\be
Y_{m,n} = (i)^{m-n} (q_2)^\ell + \hbox{``more"}, \; \; \; \; (\ell = m+n))
\ee
and hence each vector in the basis $\{Y_{m,n}\}$ ($m + n = \ell = 2s$) of the spin-$s$ subspace has a non-vanishing component along the lowest weight vector $X_{0,\ell} \sim (q_2)^\ell$.  Here, ``more" stands for the higher weight terms containing at least one $q_1$.

Under the rotation $R$, the $z$'s transform as
\be
z' = e^{ i \frac{\varphi}{2} }z, \; \; \; \; \; \bar{z}' = e^{ -i \frac{\varphi}{2} } \bar{z},
\ee
and consequently 
\be
U^{-1} Y_{m,n} U = e^{i \frac{(m-n) \varphi}{2} } Y_{m,n}.
\ee
This implies that
\begin{eqnarray}
U^{-1} \left(\sum_{m+n= \ell} \Lambda^{m,n}_0 X_{m,n} \right) U &=& U^{-1} \left(\sum_{m+n= \ell} \Xi^{m,n}_0 Y_{m,n}\right) U \hspace{4cm} \nonumber \\
& = &\sum_{m+n= \ell} \Xi^{m,n}_0 e^{i \frac{(m-n) \varphi}{2} } Y_{m,n} \nonumber \\
&=& \sum_{m+n= \ell} \Xi^{m,n}_0 e^{i \frac{(m-n) \varphi}{2} } (i)^{m-n} (q_2)^\ell + \hbox{``more"} \ .
\end{eqnarray}
We see that the coefficient of the lowest-weight basis vector $X_{0, \ell}$ in $U^{-1} \left(\sum_{m+n= \ell} \Lambda^{m,n}_0 X_{m,n} \right) U$ contains all the information on the exact symmetry $\Lambda^{\textrm{AdS}}$: its Fourier coefficients give directly the coefficients $\Xi^{m,n}_0$, or equivalently, through the change of basis (\ref{CoB}), the coefficients $ \Lambda^{m,n}_0 $ that characterize $\Lambda^{\textrm{AdS}}$.

For the spin $s$ representation, there are $(2s+1)$ Fourier exponentials in the expansion of  $\left(U^{-1} \left(\sum_{m+n= \ell} \Lambda^{m,n}_0 X_{m,n} \right) U\right)^{0,\ell}$, namely, $e^{-is \varphi}$, $e^{i(-s+1)\varphi}$, $\cdots$, $e^{i(s-1)\varphi}$, and $e^{is \varphi}$.  This exactly matches the number of coefficients $\Lambda^{m,n}_0$ ($m+n = \ell = 2s$), as it should from what we have just seen.

Note that half-integer spins have Fourier exponentials with half-integer frequencies ($e^{i \frac{\varphi}{2}}$, $e^{i \frac{3 \varphi}{2}}$, etc) and thus corresponds to anti-periodic functions, obeying  Neveu-Schwarz-like boundary conditions.

\subsubsection{Reconstructing the exact AdS symmetries from the lowest-weight components}
Recapitulating our analysis, we found that the lowest-weight components $\Lambda^{\textrm{AdS} \; 0, \ell}(\varphi)$ of the $\textrm{sl}(2,\re)$-spin $s$ generators of the exact symmetries of the AdS connection contain Fourier components with frequencies  $-s$, $-s+1$, $\cdots$, $s-1$, $s$.  From the knowledge of these lowest-weight components, we can reconstruct the complete symmetry, either by applying the route inverse to the one explained above viz. read the $ \Lambda^{m,n}_0$ from the Fourier coefficients or, alternatively and equivalently, by following a method close to the analysis  of asymptotic symmetries given in the previous section.

This method proceeds as follows. One solves the symmetry equation (\ref{LAdS}). In this case, it 
amounts to solving
\bea (\Lambda^{\textrm{AdS}})' - \frac{1}{2} [X_{11} + X_{22}, \Lambda^{\textrm{AdS}}] = 0,
\eea
starting from their lowest-weight components.  The lowest weight components of the equation give the coefficient $\Lambda^{\textrm{AdS} \; 1, \ell-1}(\varphi)$ of the symmetry generators along the basis vectors $X_{1, \ell - 1}$ in terms of the coefficients  $\Lambda^{\textrm{AdS} \; 0, \ell}(\varphi)$ , then the next equations give  $\Lambda^{\textrm{AdS} \; 2, \ell-1}(\varphi)$, etc.

The last, highest-weight component equations, which give the variation of the highest-weight component of the connection, are identically fulfilled because we have an exact symmetry.

\subsection{Exact Background Symmetries as Asymptotic Symmetries}

The above way of describing the symmetries of the AdS connection shows explicitly how the symmetries are embedded in the algebra of asymptotic symmetries, which are constructed from the lowest-weight components in exactly the same manner.  A generic asymptotic symmetry is characterized, for each spin representation, by an arbitrary periodic (integer spin) or anti-periodic (half-integer spin) function $\Lambda^{0,\ell} (\varphi)$. Only the frequencies $-s \leq k \leq s$ correspond to the AdS symmetries, and the higher Fourier components correspond to asymptotic symmetries that are not background symmetries.  Thus, for instance, in the case of the bosonic higher-spin algebra, the Fourier components $L_{-1}$, $L_0$, $L_1$ ($\textrm{sl}(2, \re)$-spin $1$ or conformal dimension $2$), $M^{(3)}_{ -2}$, $M^{(3)}_{ -1}$, $M^{(3)}_{ 0}$, $M^{(3)}_{ 1}$, $M^{(3)}_{ 2}$ ($\textrm{sl}(2, \re)$-spin $2$ or conformal dimension $3$), $M^{(4)}_{ -3}$, $M^{(4)}_{ -2}$, $M^{(4)}_{ -1}$, $M^{(4)}_{ 0}$, $M^{(4)}_{1}$, $M^{(4)}_{ 2}$, $M^{(4)}_{ 3}$ ($\textrm{sl}(2, \re)$-spin $3$ or conformal dimension $4$), etc of the bosonic $W_\infty$ algebra of [1] generate the $\textrm{hs}(2, \re)$-algebra.  The higher Fourier components (e.g., $L_2$, or $M^{(3)}_{ 3}$, or $M^{(4)}_{ -4}$ etc) generate asymptotic symmetries which are beyond $\textrm{hs}(2, \re)$.  

In the case of the $N=1$ supersymmetry, one must add $Q_{- \frac{1}{2}}$, $Q_{\frac{1}{2}}$ ($\textrm{sl}(2, \re)$-spin $1/2$ or conformal dimension $3/2$), $M^{(\frac{5}{2})}_{ -\frac{3}{2}}$, $M^{(\frac{5}{2})}_{ -\frac{1}{2}}$, $M^{(\frac{5}{2})}_{ \frac{1}{2}}$, $M^{(\frac{5}{2})}_{ \frac{3}{2}}$ ($\textrm{sl}(2, \re)$-spin $3/2$ or conformal dimension $5/2$) etc to get the superalgebra $\textrm{shs}^E(1|2, \re)$ from the corresponding super-$W_\infty$.  For $N \ge 2$ extended supersymmetry, there is an additional color index as well as the zero modes $T_0^A$ of the affine currents \footnote{Only zero modes are needed because the generators are in the spin-zero representation.}. 

There is one important point that should be stressed, however.  Even when restricted to these generators, the (super)-$W_\infty$ algebras differ from the original bulk superalgebra $\textrm{shs}^E(N | 2, \re)$ because of nonlinear terms. {\it Nevertherless, as we shall show below, the linear terms reproduce exactly the superalgebras $\textrm{shs}^E(N |2 , \re)$. Furthermore, the central charges vanish when restricted to this sector, provided the generators -- determined up to a constant -- are adjusted to be equal to zero on the AdS$_3$ connection.}  It is in that sense that the algebras $\textrm{shs}^E(N|2, \re)$ are embedded in the super-$W_\infty$ algebras. The algebra formed by the generators $\{M_n^{(\frac{j}{2}+1)} \}$ with $\vert n \vert \leq j/2$ (without the nonlinear terms)  is called the `wedge algebra' \cite{Pope:1989sr,Pope:1990be}.  Hence, one can say that, up to the nonlinear terms,  $\textrm{shs}^E(N | 2, \re)$ is embedded in $W_\infty$ as its wedge subalgebra.

The emergence of nonlinear terms is easy to understand.  Although the lowest-weight components of the asymptotic symmetries are the same for all connections asymptotic to the anti-de Sitter connection, the higher-weight ones depend on the configuration.  This is because the solution to the recursive equations determining them depends on the connection (see above and [1]).  Even if we start with a lowest-weight component that corresponds to an exact symmetry of the AdS$_3$ connection, the solution involves the deviations from the AdS$_3$ background as one works one's way up.  For that reason, the transformations of $\textrm{shs}^E(N|2, \re)$ depend on the configuration and the algebra of their generators are nonlinearly deformed.  

At this stage, one might wonder whether nonlinear deformations are the only type of deformations that can occur. The argument given below shows that this is indeed the case.

\subsection{Algebra}

Algebraically, the situation we are facing is the following: we have a set of asymptotic symmetries generated by $G_A $ ($A=\alpha, i$), which close in the Poisson bracket according to
\be
[G_A, G_B] = K_{AB} + C^C_{\; \; AB} G_C + \frac{1}{2} D^{CD}_{\; \; \; AB} G_C G_D + \cdots, 
\ee
where $K_{AB}$, $C^C_{\; \; AB}$, $D^{CD}_{\; \; \; AB}$, \dots are constants. Among these asymptotic symmetries, a subset, generated by $G_\alpha$, leaves a background (here the AdS$_3$ connection) invariant, while the others, denoted by $G_i$, do not.  The background is such that the charges-generators $G_A$ evaluated on it are zero,
\be
G_A\vert_{\hbox{Background}} = 0.
\ee
In our AdS$_3$ situation, the $G_\alpha$ are the generators associated with the lowest Fourier modes as described above, while the $G_i$ correspond to the higher Fourier modes.

Now, the transformations $\delta_\alpha F$ generated by the $G_\alpha$'s:
\be
\delta_\alpha F = [G_\alpha, F],
\ee
where $F$ is an arbitrary function of the fields, have the following properties:
\begin{enumerate}
\item When evaluated on the background, the variation $\delta_\alpha F$ vanishes,
\be \delta_\alpha F\vert_{\hbox{Background}} = 0
\ee
since the background is strictly invariant under the transformation generated by $G_\alpha$.
\item Likewise,
\be [\delta_\alpha, \delta_\beta] F\vert_{\hbox{Background}} =  f^\gamma_{\; \; \alpha \beta} \delta_\gamma F\vert_{\hbox{Background}}
\ee
where $f^\gamma_{\; \; \alpha \beta}$ are the structure constants of the background symmetry algebra.
\end{enumerate}

It follows from these two properties that 
\be
K_{A \alpha} = - K_{\alpha A} = 0
\ee
since $\delta_\alpha G_A\vert_{\hbox{Background}} = K_{A \alpha}$ (a similar argument was already used in [7,33]), and that
\be
C^\gamma_{\; \; \alpha \beta} = f^\gamma_{\; \; \alpha \beta}, \; \; \; \; \; C^i_{\; \; \alpha \beta} = 0
\ee
since $ [\delta_\alpha, \delta_\beta] F = [[G_\alpha, G_\beta],F]$, an expression that reduces to $C^C_{\; \; \alpha \beta} [G_C, F]$ on the background.

Note that the above argument says nothing about $D^{CD}_{\; \; \; \alpha \beta}$ or the higher order terms. Indeed, these can be different from zero and are actually found to be so in the present context.

{}The fact that the wedge ``subalgebra" of the W-algebra reproduces the underlying Lie algebra one starts with is actually a general result known within the Drinfeld-Sokolov reduction and demonstrated in \cite{Bowcock:1991zk} (see also \cite{Bergshoeff:1991dz,Gaberdiel:2011wb}).  We have provided here a geometrical proof within the asymptotic symmetry analysis, based on the isometries of the AdS$_3$ super-connection.

\subsection{Some Explicit Checks}

It is instructive to check explicitly the above embedding by direct computation. We shall partly do so here by computing some Poisson bracket relations chosen for their simplicity.

Translating the above relations to Fourier modes and performing the redefinition 
\be 
L_{n+m} \rightarrow L_{n+m} + \frac{k}{4}\delta_{n+m,0} \equiv \widetilde{L}_{n+m},
\ee
we obtain the following relations
\be 
\bal
~[Q_n,Q_m] &= 2k\,\delta_{n+m,0}\,(n^2-\frac{1}{4}) + 2\tilde{L}_{n+m} \\
[\widetilde{L}_n,\widetilde{L}_m] &= \frac{k}{2}\,\delta_{n+m,0}\,n(n^2-1) + (n-m)\widetilde{L}_{n+m} \\
[M^{(5/2)}_n,M^{(5/2)}_m] &= \frac{k}{18}\,\delta_{n+m,0}\,(n^2-\frac{1}{4})(n^2-\frac{9}{4}) -20 M^{(4)}_{n+m} \\
&+ \frac{1}{36}(6(n^2 + m^2) - 8nm - 9)\widetilde{L}_{n+m} + ``Q\cdot Q" \\
[M^{(3)}_n,M^{(3)}_m] &= \frac{k}{288}\,\delta_{n+m,0}\,n(n^2-1)(n^2-4) -5(n-m) M^{(4)}_{n+m} \\
&+ \frac{1}{144}(2(n^3 - m^3) - 3nm(n-m) - 8(n-m))\widetilde{L}_{n+m} \\
&+ ``Q\cdot Q" + ``\widetilde{L}\cdot \widetilde{L}" \\
[\widetilde{L}_n,M^{(j/2+1)}_m] &= (\frac{j}{2}n - m)M^{(j/2+1)}_{n+m} \quad  \textrm{($j\neq 2$)}\\
[Q_n,M^{(j/2+1)}_m] &= (j+1)M^{(\frac{j+3}{2})}_{n+m} \quad \textrm{($j$ odd $>1$)} \\
[Q_n,M^{(j/2+1)}_m] &= (n-\frac{m}{j})M^{(\frac{j+1}{2})}_{n+m} \quad \textrm{($j$ even)} .
\label{wedge00}
\eal
\ee
Here, the $`` (\hskip0.3cm) "$ on the right-hand siderefers to quadartics of the generators projected to the mode $(m+n)$. 
We observe that, as expected, when we restrict the Fourier modes to the subsector $\{M^{(j/2+1)}_n\}$ with $|n|\leq j/2$, the central charges all vanish and the linear terms on the right-hand side of the above relations only contain modes belonging to that subsector. Up to nonlinear terms, this subsector thus forms a subalgebra. 

Furthermore, one can also verify that these Poisson bracket relations are identical to those of the bulk gauge algebra $\textrm{shs}^{\textrm{E}}(N,2 \vert \re)$ once the appropriate redefinitions are made. Indeed, the analog of the above relations for $\textrm{shs}^{\textrm{E}}(N|2, \re)$ are, in the basis of the $X$'s, given by 
\be 
\bal
&[X^{(3/2)}_n,X^{(3/2)}_m] = -iX^{(2)}_{n+m} \\
&[X^{(2)}_n,X^{(2)}_m] = 2(n-m)X^{(2)}_{n+m} \\
&[M^{(5/2)}_n,M^{(5/2)}_m] = -iX^{(4)}_{n+m} + \frac{i}{2}(6(n^2 + m^2) - 8nm - 9)X^{(2)}_{n+m} \\
&[X^{(3)}_n,X^{(3)}_m] = 4(n-m) X^{(4)}_{n+m} -2(2(n^3 - m^3) - 3nm(n-m) - 8(n-m))X^{(2)}_{n+m} \\
&[X^{(2)}_n,X^{(j/2+1)}_m] = 2(\frac{j}{2}n - m)X^{(j/2+1)}_{n+m} \\
&[X^{(3/2)}_n,X^{(j/2+1)}_m] = -i X^{(\frac{j+3}{2})}_{n+m} \quad \textrm{($j$ odd)} \\
&[X^{(3/2)}_n,X^{(j/2+1)}_m] = (jn-m)X^{(\frac{j+1}{2})}_{n+m} ,
\label{wedge11}
\eal
\ee
where we have used the definition
\be 
X^{(j/2+1)}_n \equiv \frac{1}{2i}(q_1)^{2n}(q_2)^{j-2n} \quad (|n| \leq \frac{j}{2}). 
\ee

Making the redefinition
\be 
X^{(j/2+1)}_n \rightarrow \gamma^j X^{(j/2+1)}_n \equiv \tilde{X}^{(j/2+1)}_n
\ee
with
\be 
\gamma^1 = \sqrt{i}, \quad \gamma^2 = \frac{1}{2}, \quad \gamma^3 = \pm \frac{\sqrt{-i}}{6}, \quad \gamma^4 = \frac{\sqrt{-i}}{4}\gamma^3, \quad \gamma^5 = \frac{\sqrt{i}}{120}, \quad \gamma^6 = \frac{1}{720}, 
\ee
one finds the relations (\ref{wedge11}) match precisely (\ref{wedge00}).

\section{$\lambda$-Deformation}
\label{lambdaDeformed}
\setcounter{equation}{0}
So far, our consideration has been for `pure' higher-spin AdS$_3$ supergravity, viz. theory of higher-spin gauge fields only without any matter fields. One aspect of the `pure' supergravity theory was that, as mentioned before, the number of supersymmetries is arbitrary. Once matter supermultiplets are coupled, the maximal number of supersymmetries is 32 \cite{de Wit:1992up}.  Another aspect is that the matter supermultiplet has deformation-dependent masses. In this section, we shall dwell on this issue and point out that consistency of the higher-spin gauge superalgebra puts severe constraint to the mass of the matter supermutiplet. In particular, we argue that matter multiplet ought to be massless for $N > 2$ extended supersymmetries.

It is known that the gauge algebra $\textrm{hs}(2,\mathbb{R})$ can be deformed into $\textrm{hs}(2, \mathbb{R})[\lambda]$, where $\lambda$ parametrizes the ideal generated by $[C_2(\textrm{sl}(2)) - \mu \mathbb{I}]$ with
$$\mu = \frac{1}{4} (\lambda^2 - 1),$$ 
whose quotient action on the universal enveloping algebra ${\cal U}(\textrm{sl}(2))$ defines $\textrm{hs}(2, \mathbb{R})[\lambda]$ \cite{Vasiliev:1989re,Bergshoeff:1989ns}.  The undeformed case $\lambda = \frac{1}{2}$ is the case that was explicitly investigated in \cite{Henneaux:2010xg}.  However, it is trivial to verify that the general arguments of \cite{Henneaux:2010xg} apply equally well to show that the deformed case $\lambda \not=\frac{1}{2}$ also yields a nonlinear  $\textrm{W}_\infty$ algebra: the defining properties of a $\textrm{W}_\infty$-algebra spelled out in  \cite{Henneaux:2010xg} and recalled in section \ref{NonlinearW} above are fulfilled for all $\lambda$.  In fact, this algebra was conjectured in \cite{Gaberdiel:2011wb} to be the algebra constructed in \cite{FigueroaO'Farrill:1992cv} (see also \cite{Campoleoni:2011hg}).

From the viewpoint of the higher-spin algebra, the $\lambda$-deformation is realized in terms of the $\star$-commutators
\bea
[q_\alpha, q_\beta]_\star = 2i \epsilon_{\alpha \beta} [1 + (2 \lambda - 1) (-)^{N_{q}}],
\label{deformedalgebra}
\eea
where $N_{q}$ counts the number of $q$'s, so that the Klein operator $(-)^{N_{q}}$ obeys $(-)^{N_{q}} q_\alpha = - q_\alpha (-)^{N_{q}}$. It is then straightforward to check that, even after the the $\lambda$-deformation, the degree 2 basis elements $Y_{\alpha \beta}$ in (\ref{basiselement}) still form the generators of the sp$(2, \mathbb{R})$ Lie algebra (the first equation of (\ref{osp-commutator})). The hs(2$,\mathbb{R})[\lambda]$ is then constructed, schematically speaking, by extending the algebra of the $q$ variables from degree 2 monomials to all higher order polynomials. 
Notice that the deformed $q_\alpha$ obeying the $\lambda$-deformed algebra (\ref{deformedalgebra}) can be related to the original $q_\alpha$ variables obeying the undeformed Heisenberg-Weyl algebra in terms of which the definition of the $\star$-product (\ref{starproduct}) is unchanged \cite{Prokushkin:1998bq} but the algebra hs(2$,\mathbb{R})[\lambda]$ of the $\star$-commutators of polynomials of the $q$'s now depends on $\lambda$ explicitly:  the structure constants do depend on $\lambda$.

From the viewpoint of the higher-spin gravity theory, deforming the gauge algebra from $\textrm{hs}(2,\mathbb{R})$ to $\textrm{hs}(2,\mathbb{R})[\lambda]$ amounts to introducing matter fields with $\lambda$-dependent mass\footnote{Note that, at massless point, the scalar field can alternatively be described in terms of vector gauge  field.}. For example, the ideal element of the $\lambda$-deformed higher-spin algebra labels the mass  $m^2 = (\lambda^2 - 1)$ of two complex scalar matter fields introduced to the theory in addition to the massless higher-spin gauge fields \cite{Prokushkin:1998bq,Prokushkin:1998vn}. These matter scalar fields are covariantly constant with respect to the higher-spin gauge field. In a nutshell, the $\lambda$-deformation is achieved dynamically by further introducing sp$(2, \mathbb{R})$ doublet fields \cite{Prokushkin:1998bq}. Since the matter scalar fields are covariantly constant, they can take a constant vacuum expectation value,  $(2\lambda - 1)$. This expectation value then deforms the algebra satisfied by the sp$(2, \mathbb{R})$ doublet fields to the deformed sp(2,$\mathbb{R}$) algebra 
(\ref{deformedalgebra}), and facilitates the derivation of the expression of  the higher-spin algebra basis in terms of the dynamically deformed oscillators. In turn, fluctuation of the two matter scalar fields around the vacuum expectation value have the aforementioned mass spectrum.

A holographic dual conformal field theory was proposed and has passed various checks \cite{Gaberdiel:2010pz,Gaberdiel:2011wb,Gaberdiel:2011zw,Castro:2011zq,Ammon:2011ua,
Gaberdiel:2011aa}  
 though subtleties related to (de)coupling of part of the complex scalar fields is yet to be resolved \cite{Papadodimas:2011pf, Chang:2011vk}. The boundary conditions for pure gravity coupled to complex scalar fields can be relaxed to allow for generic behaviour at infinity of the scalar fields \cite{Henneaux:2002wm,Henneaux:2006hk}. The analysis should extend straightforwardly to higher-spin AdS$_3$ (super)gravity theories coupled to scalar supermultiplets. 

The $N=1$ and $N=2$ higher-spin superalgebras can also be $\lambda$-deformed \cite{Vasiliev:1999ba,Bergshoeff:1991dz}. After the deformation, the matter fields have $\lambda$-dependent mass spectrum. The matter supermultiplets are scalar fields and Majorana fermions for $N=1$, and complex scalar fields, two Dirac fermions and one vector field for $N=2$. An asymptotic analysis similar to the one given above can also be performed, yielding a deformation of the nonlinear $W$-superalgebras as symmetry algebra at AdS$_3$ infinity. A conjecture on the corresponding dual conformal field theory was investigated recently  \cite{Creutzig:2011fe} for $N=2$ (see also \cite{Candu:2012jq} for further study).

We now argue that a similar $\lambda$-deformation does not exist for $N >2$- extended supersymmetries.  Our argument relies on the fact that  the oscillator construction of the superalgebra is incompatible with the $\lambda$-deformation.  The osp$(N|2, \mathbb{R})$ generators are realized by the bilinears of the ``super-oscillators" (\ref{basiselement}). If the commutation relations of the $q$-oscillators are $\lambda$-deformed as in (\ref{deformedalgebra}), the fermionic generators of the superalgebra do not form the requisite algebra. In particular, the second through fourth (anti)commutators of (\ref{osp-commutator}) do not give rise to the osp$(N|2,\mathbb{R})$ algebra. For $N=1, 2$, the internal symmetry algebra is abelian, and the superalgebra can be made compatible with the $\lambda$-deformation by utilizing various automorphisms. This does not appear to be the case any more for $N>2$ for which the internal symmetry algebra is nonabelian. 

The clash between $N>2$-extended supersymmetry and the $\lambda$-deformation may be understood intuitively as follows. For $\lambda \ne 1/2$, the scalar matter fields are massive. The extended supersymmetry then includes in the supermultiplet other massive fields of nonzero spins. If $N >2$, the supermultiplets include massive fields with spin $s > 1$. This results in {\sl massive} higher spin gauge fields, not in the massless ones the higher-spin gauge algebra we built upon is intended to describe.

\section{Other Higher-spin Super-gauge Algebras}
\label{OtherClasses}
\setcounter{equation}{0}
We investigated so far the higher-spin superalgebra $\textrm{shs}^E(N|2, \mathbb{R})$ that extends the supergravity superalgebra $\textrm{osp}(N|2, \mathbb{R})$.  One can view this superalgebra as the quotient of the enveloping algebra of $\textrm{osp}(N|2, \mathbb{R})$ by certain ideals. This point of view can be adapted to the higher-spin superalgebras corresponding to all the other cases listed in Table 1.  

Equivalently, we have seen that a crucial ingredient used to construct $\textrm{shs}^E(N|2, \mathbb{R})$ from $\textrm{osp}(N|2, \mathbb{R})$ is that this latter superalgebra possesses a realization in terms of bosonic and fermionic oscillators. Now, the oscillator realization \cite{Gunaydin:1981yq,Bars:1982ep} (see also \cite{Tang:1984vt}) is applicable to the other classes listed in Table 1 \cite{Gunaydin:1986fe}. Therefore, higher-spin superalgebras can similarly be defined (with appropriate choice of the ideals used in the quotient of the enveloping algebra) in terms of polynomials in fermion-boson creation and destruction operators \cite{Gunaydin:1989um}. 

The oscillator construction of the regular superalgebras su$(N | 1, 1)$ and osp$( 4^* | 2N)$ is given in  \cite{Bars:1982ep,Gunaydin:1986fe}, while the one for the exceptional superalgebras  $F(4)$, $G(3)$
and D$^1(2, 1; \alpha)$ can be found in \cite{DellaSelva:1992fp}.  One goes from these finite superalgebras to the corresponding higher spin algebras by relaxing the condition on the polynomial degree.

We do not intend to complete the construction of the other higher-spin superalgebras in this section but rather want to point out that, in each case, one gets a nonlinear super-W$_\infty$  algebra as the algebra of asymptotic symmetries at infinity for the associated higher-spin AdS$_3$ supergravity theory.  This is because, in all cases, the algebra contains the gravitational sl$(2,\re)$ subalgebra under which it decomposes as a  direct sum of finite-dimensional representations of increasing sl$(2,\re)$-spins $s$.  The AdS$_3$ boundary conditions keep only the fields associated with the highest-weight states. These become  asymptotically the generators of conformal spin $s +1$. 

All these general features are independent of the structure of the underlying superalgebra and follow the pattern described for $\textrm{osp}(N|2, \mathbb{R})$.  Of course, the precise representations that occur and the detailed form of the Poisson bracket relations (besides the brackets with the Virasoro generators and the supercharges) depend on the particular superalgebra being considered (as well as on its possible deformations in what concerns the brackets).

In the Appendix D, we  illustrate how the construction of other higher-spin superalgebras proceeds for the case of the higher-spin superalgebra shs$(N|1,1)$ built upon su$(1,1|N)$, and briefly comment on the starting setups for the other classes.

One final comment pertains to rigidity.  One may argue again that the $N>2$ higher-spin superalgebras constructed out of the different superalgebras do not admit the $\lambda$-deformation.  This is because, just as for the osp$(N|2, \mathbb{R})$ class, the fermionic bilinear basis elements do not satisfy the 
requisite (anti)commutation relations. This indicates that $N>2$ higher-spin superalgebras has rigidity against the $\lambda$-deformation. For maximally supersymmetric higher-spin superalgebras, this is as one should expect since all the fields should form a single supermultiplet.

\section{String Theory Realization}
\label{StringTReal}

It is of interest to identify string theory setups where the higher-spin supergravity theories might arise, in particular, with $(M, N)$-extended supersymmetries and gauge superalgebras studied in the previous sections. Here, furthering the discussions in \cite{Henneaux:2010xg}, we point out that a family of such setups could arise as the near-horizon dynamics of macroscopic fundamental strings in compactified heterotic or type II string theories.

Consider a stack of $n$ many infinitely extended fundamental strings, preserving half of the 16 or 32 spacetime supersymmetries. The supergravity solution of the macroscopic strings~\cite{Dabholkar:1990yf} turns out to exhibit that the metric at the center is singular with divergent spacetime curvature invariants and that the string coupling at the center is zero. Since the solution is singular, higher-derivative corrections to the leading-order supergravity action are important. They are generated in string worldsheet perturbation theory. With these string worldsheet corrections, the singular geometry might be resolved to a regular one and a holographic dual of the macroscopic fundamental string can be obtained.

Such a possibility was investigated in~\cite{Dabholkar:2007gp,Lapan:2007jx,Kraus:2007vu,Johnson:2007du}, and evidence was accumulated that the string-corrected near-horizon geometry indeed becomes nonsingular into AdS$_3 \times X$, where the geometry of $X$ depends on the details of the compactification. The near-horizon geometry also doubles the conserved supersymmetries. In a nutshell, the higher derivative corrections lead to a `stretched horizon', where the horizon size is of the order of the string scale. By varying the dimension of the compactified tori, it turns out all seven classes of the spacetime isometry superalgebra in Table 1 can be realized.

The above result was obtained in the limit where the number of  fundamental strings and the chiral momenta modulated along the strings is large. If these charges are taken to be smaller and become of order one, the curvature radii of AdS$_3$ and $X$ would diminish to zero (in units of the string scale). Therefore,
\bea
T_{\rm st} R_{\rm ads}^2 = {R^2_{\rm ads} \over \ell_{\rm st}^2} \ll 1,
\eea
and the string tension (measured in units of the curvature scale of the near-horizon geometry) becomes zero. In fact, the AdS$_3$ string theory undergoes a phase transition as the tension is reduced across the critical point $T_{\rm st} R_{\rm ads}^2 = 1$: both the SL$(2, \mathbb{C})$ invariant vacuum and the BTZ black hole states of the spacetime CFTs become non-normalizable
\cite{Giveon:2005mi}, and the density of high-energy states changes discontinuously \cite{Nakayama:2005pk}.

Below the critical point, a typical AdS$_3$ string state is a long string that extends to the boundary and becomes non-interacting. The tensionless string would lead to massless higher-spin fields in the AdS$_3$ bulk, and, by our analysis, the super-W$_\infty$ algebra will emerge as the asymptotic symmetry.

%
\begin{table}
\label{Table2}
\begin{centering}
\begin{tabular}{|c|c|c|c|}
\hline
tori & isometry & R-symmetry & superalgebra \\
\hline
$\mathbb{T}^0$ & spin(8) & spin(8) &
osp$(8|2, \mathbb{R})$ \\
$\mathbb{T}^1$ & spin(7) & spin(7) &
F(4) \\
$\mathbb{T}^2$ & spin(6) & spin(6)$\times$spin(2) &
su$(1,1|4)$ \\
$\mathbb{T}^3$ & spin(5) & spin(5)$\times$spin(3) &
osp$(4^*|4)$ \\
$\mathbb{T}^4$ & spin(4) & spin(4)$\times$spin(4) &
(D$^1(2,1;\alpha))^2$ \\
$\mathbb{T}^5$ & spin(3) & spin(3)$\times$spin(5) &
osp$(4^*|4)$ \\
\hline
\end{tabular}
\caption{\footnotesize{The anticipated global symmetry of the near-horizon in various compactifications of macroscopic superstring. The first column refers to the tori $\mathbb{T}^d$ transverse to $\mathbb{R}^{1,1} \times \mathbb{R}^{8-d}$, the second column to the near-horizon isometry, the third column to the R-symmetry of the supercharges and the last colum the relevant superalgebra for 16 supercharges of the half-BPS macroscopic fundamenatl string.} }

\end{centering}
\end{table}

%


%
For the compactification on $\mathbb{T}^{d} (0 \le d \le 5)$, the near-horizon isometry, R-symmetry and anticipated superalgebra are given in Table 2 (see \cite{Dabholkar:2007gp}).
Comparing Table 1 and Table 2, we see that all classes except those based on $G(3)$ are possibly realizable as the higher-spin superalgebras. Furthermore, for the superalgebras in the infinite families, the near-horizon geometry with lower supersymmetries can be obtained either by orbifolding the toroidal directions or by wrapping NS5-branes or heterotic 5-branes.

It is also interesting to note that the missing $G(3)$ class is actually present in all possible maximally supersymmetric three-dimensional gauged supergravity theories. The complete classification was accomplished in \cite{Nicolai:2000sc, deWit:2004yr}, 
which we list in Table 3.  A striking feature is that these theories are in general heterotic -- left and right supersymmetries are different. It would be a very interesting question left for further study to investigate whether these left-right asymmetric theories can be embedded to string theory compactifications. 
%
%
\begin{table}
\label{Table3}
\begin{centering}
\begin{tabular}{|c|c|c|}
\hline
gauge algebra & $(n_L, n_R)$ & ground-state isometry \\
\hline
so(8)$\times$so(8) & (8,8) &
osp$(8|2, \mathbb{R}) \times$ osp $(8|2, \mathbb{R})$ \\
so(7,1)$\times$so(7,1) & (8,8) &
F(4)$\times$F(4) \\
so(6,2)$\times$so(6,2) & (8,8) &
su$(1,1|4) \times$su$(1,1|4)$ \\
so(5,3)$\times$so(5,3) & (8,8) &
osp$(4^*|4) \times$ osp$(4^*|4)$ \\
so(4,4)$\times$so(4,4)& (8,8) &
$\mathbb{R}^{2,1}$\\
G$_{2(2)} \times$ F$_{4(4)}$ & (4, 12) &
D$^1(2, 1;-{2\over 3}) \times$ osp($4^*|6)$ \\
G$_2 \times$ F$_{4(-20)}$ & (7,9) & G(3)$\times$osp$(9|2,\mathbb{R})$ \\
E$_{6(6)}\times$ sl(3) & (16,0) & osp$(4^*|8) \times$ su(1,1) \\
E$_{6(2)}\times$su(2,1) & (12,4)& su$(6|1,1)\times$D$^1(2,1;-{1 \over 2})$ \\
E$_{6(-14)}\times$su(3) & (10,6) & osp$(10|2, \mathbb{R})\times$su$(3|1,1)$ \\
E$_{7(7)} \times$sl(2) & (16,0) & su$(8|1,1) \times $ su$(1,1)$ \\
E$_{7(-5)} \times$ su(2) & (12,4) & osp$(12|2, \mathbb{R}) \times$D$^1(2,1; -{1 \over 3})$\\
E$_{8(8)}$ & (16,0) & osp$(16|2,\mathbb{R})\times$ su(1,1)\\
\hline
\end{tabular}
\caption{\footnotesize{The maximally supersymmetric gauged supergravity theories with semi-simple gauge algebra.} }
\end{centering}
\end{table}

The AdS/CFT correspondence then asserts that there ought to be holographic dual CFTs (spacetime CFTs) to each of these near-horizon isometries and higher-spin supergravity theories. By a chain of S- and T-dualities, the dual CFTs can be defined in terms of the M(atrix) string theories for Type II \cite{Dijkgraaf:1997vv} and heterotic \cite{Rey:1997hj} strings.

As mentioned above, a proposal for the holographic dual spacetime CFT was recently put forward for $N=0$ and $2$ \cite{Gaberdiel:2010pz,Creutzig:2011fe,Candu:2012jq}.  Let us recapitulate this for the bosonic case first.

One starts with the observation that in three dimensions the higher spin (super)algebras can be truncated up to a finite spin $n$. Correspondingly, the gauge algebra hs$(2, \mathbb{R})$ is truncated to sl$(n, \mathbb{R})$ (which corresponds  to the particular value $\lambda = N$ of hs$(2, \mathbb{R})[\lambda]$), and the asymptotic symmetry algebra is the W$_n$-algebra. On the other hand, it has been known that the coset CFT
\bea
{G \over H}(n,k) = {SU(n)_k \times SU(n)_1 \over SU(n)_{k+1}}
\label{bosonic-coset}
\eea
is the minimal model that has the extended symmetry of W$_n$ algebra. Based on this symmetry consideration, it was proposed that the higher-spin gravity theory is the holographic dual to the coset CFT (\ref{bosonic-coset}) \cite{Gaberdiel:2010pz}. The CFT has two coupling parameters $n, k$ and the proposed correspondence was tested in the `shifted' `t Hooft limit~\footnote{The `t Hooft coupling is called `shifted' since the level $k$ is taken shifted by $N$.}
\bea
n, k \rightarrow \infty \qquad \mbox{and} \qquad \lambda \equiv {n \over k + n} = \mbox{finite}.
\eea
The proposal actually contains two complex matter fields coupled to the higher-spin gravity theory. The mass can be varied by deforming the higher- spin algebra hs$(2, \mathbb{R})[\lambda]$.

For the ${N}=2$ case, the dual CFT has the $N=2$-extended super-${\cal W}_{n}$ symmetry. It has also been known that the minimal model of the spacetime CFTs with this symmetry algebra is the Kazama-Suzuki coset models over $\mathbb{CP}^n$ \cite{Kazama:1988qp}:
\bea
{G \over H} = {SU(n+1)_k \times SO(2n)_1 \over SU(n)_{k+1} \times U(1)_{n(n+1)(n + k + 1)}}.
\eea
One again has two coupling parameters $n, k$, and the shifted `t Hooft limit is also understood. In fact, the Kazama-Suzuki coset models were originally constructed by endowing the GKO coset models with ${ N}=2$-extended superconformal symmetry.

This line of reasoning suggests that the spacetime CFT dual to $N$-extended higher-spin supergravity is obtainable by endowing the GKO coset model with $N$-exended superconformal symmetry. For $N=4$, such a construction was already studied in \cite{VanProeyen:1989me,Sevrin:1989ce}. These authors found that the corresponding coset models must be of the type
\bea
{G \over H} =
W \otimes \mbox{SU}(2) \times \mbox{U}(1),
\label{n=4coset}
\eea
where $W$ refers to a so-called Wolf symmetric space.

Recall that a Wolf space is a quaternionic symmetric space. It is known that for every semisimple Lie algebra, there exists a corresponding Wolf space. For example, for $A_n$, $B_n$ and $D_n$ and $C_n$, the associated Wolf spaces are
\bea
&& W_u(n) = {SU(n) \over S(U(n-2) \times U(2))} \nonumber \\
&& W_o(n) = {SO(n+4) \over SO(n) \times SO(4)}
\nonumber \\
&& W_p (n) = {Sp(n) \over Sp(n-1) \times Sp(1)},
\eea
respectively, and have dimensions $4(n-2)$.

The coset models (\ref{n=4coset}) are the CFTs with the standard, linear $N=4$ superconformal algebra. On other other hand, if the U(1) current in (\ref{n=4coset}) is further quotiented out, the resulting odd-dimensional coset models are the CFTs with the nonlinear $N=4$ superconformal field theories of Bershadsky-Knizhnik type \cite{Sevrin:1989ce}.

For $N >4$, little is known about the minimal coset models that can be identified as the candidate spacetime CFTs. It would be interesting to expand the GKO coset construction to $N>4$ superconformal symmetries and, if such cosets exist,  construct candidate spacetime CFTs.

\section{Discussion}

In this paper, we studied the ${\cal N}=(N,M)$-extended higher spin AdS$_3$ supergravity theory and showed that the asymptotic symmetry is an extended nonlinear super-W$_\infty$ algebra. This proof is built upon the analysis of \cite{Henneaux:2010xg} for bosonic higher-spin gravity theory and the analysis of \cite{Henneaux:1999ib} for ${\cal N}=(N,M)$-extended AdS$_3$ supergravity theories.

One striking feature of the asymptotic algebra is that the classical central charge $c = 3 \ell / 2 G_N$ is independent of the number of higher-spin fields and the amount of extended supersymmetries. Technically, this follows from the fact that the gravitational $\textrm{sl}(2, \mathbb{R})$ is always the same so that the computation of the classical central charge is unaffected by the presence of the extra fields.

Another feature is that the nonlinear deformation of the super-W$_\infty$ algebra is intimately intertwined with the quadratic deformation of the extended superconformal algebras, in the sense that  the nonlinearities of the super-W$_\infty$ algebras contain the nonlinearities of the quadratically-deformed extended superconformal algebras.

We pointed out that the near-horizon geometry of macroscopic fundamental strings is one possible string theory realization of the higher-spin AdS$_3$ supergravity theory. We initiated the identification of the holographic dual spacetime CFTs for the $N=4$ case, but further study of these CFTs and of the $N>4$ case is relegated to the future.

Among the seven classes of the identified higher-spin AdS$_3$ supergravity theories, we still do not have candidate string theory realizations for the classes based on the G$(3)$ superalgebras. On the other hand,  G$(3)$ appears prominently in the left-right asymmetric gauged supergravity with maximal supersymmetries. It would be interesting to investigate whether such theories can be realized from some string theory compactifications.

So far, we restricted our study of higher-spin dynamics to the classical regime $k \rightarrow \infty$. It would be very interesting to extend the study to the quantum regime $k \sim {\cal O}(1)$.
The BRST quantization of a closely related setup was studied sometime ago \cite{Bars:2001ma}, and it should be readily applicable to the higher-spin AdS$_3$ (super)gravity theories. This way, we would be able to open the prospect for the complete understanding of strong curvature and strong quantum aspects of gravitational dynamics via the higher-spin gravity theory.

Finally, extending what we already emphasized in \cite{Henneaux:2010xg}, it would be highly rewarding to find and classify new black holes in the higher-spin AdS$_3$ supergravity theories constructed in this paper (see \cite{Gutperle:2011kf,Ammon:2011nk,Castro:2011fm,Tan:2011tj,Gaberdiel:2012yb} for black hole solutions of the bosonic theories with $W_N$ symmetry). Perhaps, the black hole might not be that special and appear as an emergent geometry, as hinted in \cite{Rey:2006bz}.

\section*{Acknowledgements}
We thank Xavier Bekaert, Michael Douglas, Hermann Nicolai, Adam Schwimmer, Yuji Sugawara, Misha Vasiliev for useful discussions and Murat Gunaydin for comments on a first version of this paper. S.J.R. thanks Jim Morgan and Mike Douglas for warm hospitality during his stays to the Simons Center for Geometry and Physics.  M. H. thanks the Institute for Advanced Study (Princeton) for warm hospitality while this work was being completed. He also gratefully acknowledges support from the Alexander von Humboldt Foundation through a Humboldt Research Award and support from the ERC through the ``SyDuGraM" Advanced Grant. The work of M. H. and  G.L.G was partially supported by IISN
- Belgium (conventions 4.4511.06 and 4.4514.08), by the Belgian Federal Science Policy Office through the Interuniversity Attraction Pole P6/11 and by the ``Communaut\'e Fran\c{c}aise de Belgique" through the ARC program. The work of J.S.P and S.J.R was supported in part by the National Research Foundation of Korea grants 2005-009-3843, 2009-008-0372 and 2010-220-C00003.

\appendix
\section{Conventions and Notations}
\label{A}
$\bullet$ \underline{$\textrm{sl}(2, \mathbb{R})$} \hfill\break
A commuting spinor $q$ of $\textrm{sl}(2, \mathbb{R})$ is a two-component, real-valued column vector
\bea
q \equiv ( q^\alpha) = \left( \begin{array}{c} q^1 \\ q^2 \end{array} \right) \qquad (\alpha = 1, 2).
\eea
The spinor indices are raised and lowed with the spinor metric
\bea
(\epsilon^{\alpha \beta}) = (\epsilon_{\alpha \beta}) = \left( \begin{array}{cc} 0 & + 1 \\ - 1 & 0 \end{array}
\right), \qquad (\alpha, \beta = 1, 2)
\eea
in the North-West/South-East convention:
\bea
A_\alpha = A^\beta \epsilon_{\beta \alpha}, \qquad A^\alpha = \epsilon^{\alpha \beta} A_\beta.
\eea

$\bullet$ \underline{$\textrm{AdS}_3$} \hfill\break
Denote AdS$_3$ radius as $\ell$. We adopt the global coordinates of AdS$_3$:
\bea
(x) = (x^0, x^1, x^2) = (t, \ell \theta, r) .
\eea
in which the metric reads
\be
ds^2 = - \left(1 + \left(\frac{x^2}{\ell}\right)^2\right) (dx^0)^2 +\left(1 + \left(\frac{x^2}{\ell}\right)^2\right)^{-1} (dx^2)^2 +  \left(\frac{x^2}{\ell}\right)^2 (d x^1)^2
\ee
To leading order at infinity, the ``$1$" is negligible and one can replace asymptotically the metric by that of the zero mass black hole \cite{Banados:1992gq},
\be
ds^2 = - \left(\frac{x^2}{\ell}\right)^2 (dx^0)^2 + \left(\frac{x^2}{\ell}\right)^{-2} (dx^2)^2 +  \left(\frac{x^2}{\ell}\right)^2 (dx^1)^2
\ee

The light-cone coordinates are defined by
\bea
(x) = (x^\pm, x^2) = ( t \pm \ell \theta, r).
\eea

\section{Matrix realization of osp$(N|2, \mathbb{R})$ superalgebra}
We collect useful result for matrix realization of osp$(N|2, \mathbb{R})$ superalgebra.

\subsection{The non-extended case}

The orthosymplectic $\textrm{osp}(1,2|\mathbb{R})$ superalgebra can be realized as the real vector space of even (grading-preserving) $3\times 3$ supermatrices acting on $1$ commuting real Grassmann variable $x$ and $2$ anticommuting real Grassmann variables $\theta^1$ and $\theta^2$ and which preserve the quadratic form
\be 
x^2 + 2i\theta^1\theta^12 = x^2 + i\epsilon_{\alpha\beta}\theta^\alpha\theta^\beta 
\ee 
as well as the real character of the coordinates, with the usual Lie bracket 
\be 
[\Gamma,\Gamma'] \equiv \Gamma\Gamma' - \Gamma'\Gamma ,
\ee
where the multiplication is the matrix multiplication. Such supermatrices have the form
\be 
\bmat
0 & i\mu & -i\lambda \\
\lambda & a & b \\
\mu & c & -a 
\emat
\ee
with $a,\, b,\, c$ real and commuting and $\lambda,\, \mu$ real and anticommuting. We identify the generators:
\be 
\label{osprepresentation}
\bal
H &\equiv \bmat 0 & 0 & 0 \\ 0 & 1 & 0 \\ 0 & 0 & -1 \emat , \qquad E \equiv \bmat 0 & 0 & 0 \\ 0 & 0 & 1 \\ 0 & 0 & 0 \emat , \qquad F \equiv \bmat 0 & 0 & 0 \\ 0 & 0 & 0 \\ 0 & 1 & 0 \emat , \\
R^+ &\equiv \bmat 0 & i & 0 \\ 0 & 0 & 0 \\ 1 & 0 & 0 \emat , \qquad R^- \equiv \bmat 0 & 0 & -i \\ 1 & 0 & 0 \\ 0 & 0 & 0 \emat , 
\eal
\ee
according to which we find the supercommutators
\begin{alignat*}{3}
[H,E] &= 2E, {}&{} [H,F] &= -2F, {}&{} [E,F] &=H, \\
[H,R^+] &= R^+, {}&{} [E,R^+] &= 0, {}&{} [F,R^+] &=R^-, \\
[H,R^-] &= -R^-, {}&{} [E,R^-] &= R^+, {}&{} [F,R^-] &=0, \\
\{R^+,R^+\} &= -2iE \quad {}&{} \{R^-,R^-\} &= 2iF \quad {}&{} \{R^+,R^-\} &= iH  
\end{alignat*}
where the supercommutator is defined in the usual way
\be
[\Gamma,\Gamma'\} \equiv \Gamma\Gamma' - (-)^{\pi_\Gamma\pi_{\Gamma'}}\Gamma'\Gamma .
\ee
The supertrace and scalar product are defined as
\be 
\str(\Gamma) \equiv \Gamma_{11} - \Tr(\Gamma_{\textrm{sp}(2)}) = \Gamma_{11} - \Gamma_{22} - \Gamma_{33} = - \Gamma_{22} - \Gamma_{33},
\ee
\be
(\Gamma,\Gamma') \equiv \str(\Gamma\Gamma') , 
\ee
where $\Gamma_{\textrm{sp}(2)}$ is the submatrix generated by $E$, $F$ and $H$ (``spacetime'' algebra), and there is no internal algebra because $N=1$. In our representation, the fermionic sector is thus encoded in the $F_{1a}$ and $F_{a1}$ components of the matrices and the $\textrm{sp}(2|\mathbb{R})$ subalgebra of $\textrm{osp}(1,2|\mathbb{R})$ thus lies in the $F_{ab}$ components, with $a,b = 1,2$. 

\subsection{The extended case}

The orthosymplectic $\textrm{osp}(N,2|\mathbb{R})$ superalgebra can be realized as the real vector space of even (grading-preserving) $(N+2)\times(N+2)$ supermatrices acting on $N$ commuting real Grassmann variables $x^i$ and $2$ anticommuting real Grassmann variables $\theta^1$ and $\theta^2$ and which preserve the quadratic form
\be 
\sum_{i=1}^{N}(x^i)^2 + 2i\theta^1\theta^2 = \delta_{ij}x^ix^j + i\epsilon_{\alpha\beta}\theta^\alpha\theta^\beta 
\ee 
as well as the real character of the coordinates, with the usual Lie bracket 
\be 
[\Gamma,\Gamma'] \equiv \Gamma\Gamma' - \Gamma'\Gamma ,
\ee
where the multiplication is the matrix multiplication. Such supermatrices have the form
\be 
\bmat
~ & i\mu_1 & -i\lambda_1 \\
O_{ij} & \vdots & \vdots \\
~ & i\mu_N & -i\lambda_N \\ 
\lambda_1\;\cdots\;\lambda_N & a & b \\
\mu_1\;\cdots\;\mu_N & c & -a 
\emat
\ee
with $O_{ij} = -O_{ji}$ and $a,\, b,\, c$ real and commuting and $\lambda_i,\, \mu_i$ real and anticommuting. We identify the generators:
\be 
\label{eosprepresentation}
\bal
H &\equiv \bmat ~ & 0 & 0 \\ 0 & \vdots & \vdots \\ ~ & 0 & 0 \\ 0\;\cdots\;0 & 1 & 0 \\ 0\;\cdots\;0 & 0 & -1 \emat , \qquad E \equiv \bmat ~ & 0 & 0 \\ 0 & \vdots & \vdots \\ ~ & 0 & 0 \\ 0\;\cdots\;0 & 0 & 1 \\ 0\;\cdots\;0 & 0 & 0 \emat , \qquad F \equiv \bmat ~ & 0 & 0 \\ 0 & \vdots & \vdots \\ ~ & 0 & 0 \\ 0\;\cdots\;0 & 0 & 0 \\ 0\;\cdots\;0 & 1 & 0 \emat , \\
R^+_i &\equiv \bmat ~ & 0 & 0 \\ ~ & \vdots & \vdots \\ 0 & 0 & -i \\ ~ & \vdots & \vdots \\ ~ & 0 & 0 \\ 0\;\cdots\;1\;\cdots\;0 & 0 & 0 \\ 0\;\cdots\;0\;\cdots\;0 & 0 & 0 \emat , \quad R^-_i \equiv \bmat ~ & 0 & 0 \\ ~ & \vdots & \vdots \\ 0 & i & 0 \\ ~ & \vdots & \vdots \\ ~ & 0 & 0 \\ 0\;\cdots\;0\;\cdots\;0 & 0 & 0 \\ 0\;\cdots\;1\;\cdots\;0 & 0 & 0 \emat ,\\
J_{ij} &\equiv \bmat 0 \qquad 1 & 0 & 0 \\ \ddots & \vdots & \vdots \\ \bar{1}\hspace*{24pt}0 & 0 & 0 \\ 0\;\cdots\;0 & 0 & 0 \\ 0\;\cdots\;0 & 0 & 0 \emat ,
\eal
\ee
where in $R^+_i$ and $R^-_i$ (odd generators) the $i$ factors sit in the $i$-th line and the $1$ factors in the $i$-th column, and in $J_{ij}$ the $1$ (resp. $-1$) factors sit in the position $(i,j)$ (resp. $(j,i)$). We find the supercommutators
\begin{alignat*}{3}
[H,E] &= 2E, {}&{} [H,F] &= -2F, {}&{} [E,F] &=H, \\
[H,R^+_i] &= R^+_i, {}&{} [E,R^+_i] &= 0, {}&{} [F,R^+_i] &=R^-_i, \\
[H,R^-_i] &= -R^-_i, {}&{} [E,R^-_i] &= R^+_i, {}&{} [F,R^-_i] &=0, \\
i\{R^+_i,R^+_j\} &= 2\delta_{ij}E \quad {}&{} i\{R^-_i,R^-_j\} &= -2\delta_{ij}F \quad {}&{} i\{R^+_i,R^-_j\} &= J_{ij} - \delta_{ij}H \\
[J_{ij},E] &= 0, {}&{} [J_{ij},F] &= 0, {}&{} [J_{ij},H] &=0
\end{alignat*}
\vspace*{-22pt}
\be 
[J_{ij},R^+_k] = \delta_{jk}R^+_i - \delta_{ik}R^+_j \qquad [J_{ij},R^-_k] = \delta_{jk}R^-_i - \delta_{ik}R^-_j  
\ee
\vspace*{-16pt}
\be 
[J_{ij},J_{kl}] = \delta_{jk}J_{il} + \delta_{il}J_{jk} - \delta_{ik}J_{jl} - \delta_{jl}J_{ik},
\ee
where the supercommutator is defined in the usual way
\be
[\Gamma,\Gamma'\} \equiv \Gamma\Gamma' - (-)^{\pi_\Gamma\pi_{\Gamma'}}\Gamma'\Gamma .
\ee
The supertrace and scalar product are defined as
\be 
\str(\Gamma) \equiv \Tr(\Gamma_{\textrm{so}(N)}) - \Tr(\Gamma_{\textrm{sp}(2)}),
\ee
\be
(\Gamma,\Gamma') \equiv \str(\Gamma\Gamma') , 
\ee
where $\Gamma_{\textrm{so}(N)}$ is the submatrix of $\Gamma$ generated by the $J_{ij}$ basis elements (internal algebra) and $\Gamma_{\textrm{sp}(2)}$ is the submatrix generated by $E$, $F$ and $H$ (``spacetime'' algebra).

\section{Some useful relations for the $N=1$ case}
\label{AppB}
\setcounter{equation}{0}

The $\textrm{sh}(1,1\vert \mathbb{R})$ algebra can be consistently truncated to spin $\le 2$. In the basis
\begin{equation}
X_{11} = \frac{1}{4  i} \left( q^1\right)^2, \; \; \; X_{12} =  \frac{1}{2  i} q^1 q^2, \; \; \; X_{22} = \frac{1}{4  i}  \left( q^2\right)^2,\; \; \; X_1 = \frac{1}{2  i} q^1, \; \; \; X_2 = \frac{1}{2  i} q^2,
\ee
the nonzero super-commutators are
\bea
&& [X_{11}, X_{22}]_* = X_{12}, \quad [X_{12}, X_{11}]_* = - 2 X_{11}, \quad [X_{12}, X_{22}]_* = + 2 X_{22} \nonumber \\
&& \{ X_1, X_1 \}_* = - 2i X_{11}, \quad \{X_2, X_2 \}_* = - 2i X_{22}, \quad \{X_1, X_2 \}_* = - i X_{12} \nonumber \\
&& [X_1, X_{12} ]_* = X_1, [X_1, X_{22}]_* = X_2, \quad [X_2, X_{11}]_* = - X_1, \quad [X_2, X_{12}]_* = - X_2.
\eea
This is nothing but the Lie superalgebra $\textrm{osp}(1, 2 \vert \mathbb{R})$: in Chevalley basis, $(E, F, H) = (X_{11}, -X_{22}, -X_{12})$ and $(R_+, R_-) = (X_1, X_2)$.

To have a nonzero scalar product between two basis elements,  the total number of $1$-indices must match the total number of $2$-indices. Physically, this implies that the invariant subspaces of different $\textrm{sl}(2, \mathbb{R})$-spin representations  are  orthogonal.  A few examples illustrate this:
\bea
&& (X_1, X_2) = -2i \nonumber \\
&& (X_{11}, X_{22}) = 1, \qquad (X_{12}, X_{12}) = -2, \cdots \nonumber \\
&& (X_{111}, X_{222}) = {i \over 3}, \qquad (X_{112}, X_{122}) =  -i, \cdots \nonumber \\
&& (X_{1111}, X_{2222}) = -{1 \over 12}, \quad (X_{1112}, X_{1222}) = {1 \over 3}, \quad
(X_{1122}, X_{1122}) = -{1 \over 2}, \cdots.
\eea
Likewise,
\bea
(X_{(s,0)}, X_{(0, s)}) = - 2{i^s \over s!}.
\eea
This inner product structure indicates that the $\textrm{shs}(1,1 \vert \mathbb{R})$ superalgebra can be thought of as a stack of mutually orthogonal $\textrm{sl}(2, \mathbb{R})$-spin $s$ layers.

Of course, different spin layers mix  through the $\star$-bracket, as can be seen for instance from the following (anti)commutators,
\bea
\{ X_{(m, 0)}, X_{(0, n)} \}_* = - i \sum_{\ell = 0}^{u(m,n)} (-)^\ell {1 \over (2 \ell)!} X_{(m-2 \ell, n - 2 \ell)}
\eea
for $m, n$ both odd integers and
\bea
[X_{(m,0)}, X_{(0, n)}]_* = \sum_{\ell = 0}^{u(m,n)} (-)^\ell {1 \over (2 \ell + 1)!} X_{(m - 2 \ell - 1, n - 2 \ell - 1)}
\eea
for $m$ or $n$ even integer, where $u(m, n) \equiv (r - 2^{1 - \pi(r)})/2$ and $r \equiv \mbox{min}(m, n)$.  Here, $\pi(r) = 0$ for $r$ even and $\pi(r) = 1$ for $r$ odd.

\section{Construction of other higher-spin superalgebras}
\subsection{su$(1, 1|N)$ ($N \ge 2$)}

The su$(1,1|N)$ superalgebra for $N \ne 2$ has the Jordan decomposition structure:
\bea
\mbox{su}(1,1|N) = {\cal L}^- \oplus {\cal L}^0 \oplus {\cal L}^+
\eea
with respect to the maximal compact sub-superalgebra~\footnote{The choice of the maximal compact sub-superalgebra is not unique. This is reflected in the fact that here are various possible choices of $L = 0, \cdots, N$ in (\ref{maximalsub}).}
\bea
{\cal L}^0 = \mbox{su}(1 | L) \oplus \mbox{su} (1 | N-L) \oplus \mbox{u}(1)
\label{maximalsub}
\eea
and
\bea
[{\cal L}^0, {\cal L}^\pm] = {\cal L}^\pm, \qquad
[{\cal L}^-, {\cal L}^+] = {\cal L}^0,
\qquad
[{\cal L}^\pm, {\cal L}^\pm] = 0.
\eea

We introduce two bosonic variables $q_\alpha$ and $L + (N-L)$ fermionic variables $\psi_i$ that transform covariantly under the su$(1|L)$ and su$(1|N-L)$ sub-superalgebras. These variables are complex-valued and so their complex conjugates are contravariant vectors $(q_\alpha)^\dagger = q^\alpha, (\psi_i)^\dagger = \psi^i$. We can build two super-variables
$\xi_A, \eta_M$:
\bea
u \equiv \{\xi_A, \eta_M \}: \qquad \xi_A = (q_1, \psi_1 \cdots, \psi_L), \qquad
\eta_M = (q_2, \psi_{L+1}, \cdots, \psi_N)
\eea
and their conjugates $\xi^A, \eta^M$. In the space of functions $f(z)$ of the commuting variables $q_\alpha$, $q^\alpha$ and anti-commuting variables $\psi_i$, $\psi^i$ ($(z)= (u, u^\dagger)$), we also
define the $\star$-product:
\bea
(f \star g)(z'') \equiv \exp \left( \left[{\partial \over \partial q_\alpha}{\partial \over \partial q'^\alpha} - {\partial \over \partial q^\alpha} {\partial \over \partial q'_\alpha} \right] + {\partial \over \partial \psi_i} {\partial \over \partial \psi'^i} + {\partial \over \partial \psi^i} {\partial \over \partial \psi'_i } \right)
f(z) g(z') \Big\vert_{z = z' = z''}. \ \ \ \
\label{su-starproduct}
\eea

Then, the $\star$-(anti)commutators realize the Weyl, respectively, Clifford algebra for the bosonic and fermionic variables:
\bea
&& \left[\ q_\alpha \ , \ q^\beta \ \right]_\star = \delta^\beta_\alpha \qquad
\alpha, \beta = 1, 2
\nonumber \\
&& \left\{ \psi_i, \psi^j \right\}_\star = \delta^j_i \qquad i, j = 1, \cdots, N \nonumber \\
&& \left[ \ q_\alpha \ , \ q_\beta \ \right]_\star = \left[\ q^\alpha \ , \ q^\beta \ \right]_\star = \left\{ \psi_i, \psi_j \right\}_\star \left\{\psi^i, \psi^j \right\}_\star = 0 .
\eea

We now introduce a ``$\mathfrak{N}$-degree" that counts the super-variables, giving weight $+1$ to $\xi^A$ and $\eta_M$, and weight $-1$ to $\xi_A$ and $\eta^M$.  More precisely,
\be 
\mathfrak{N}(\xi^A)  = 1 , \; \; \mathfrak{N}(\xi_A)  = - 1 , \; \; \mathfrak{N}(\eta^M) =- 1 , \; \; \mathfrak{N}(\eta_M) = 1 , 
\ee
so that if we call $m_A$ (respectively $\overline{m}_M$) the degree of a monomial $P$ in the creation variables $\xi^A$ (respectively $\eta^M$) and $n_A$ (respectively $\overline{n}_M$) its degree in the destruction variables $\eta_M$, $ P = (\xi^1)^{m_1} (\xi^2)^{m_2} \cdots (\eta^1)^{\bar{m}_1} \cdots (\xi_1)^{n_1} \cdots (\eta_1)^{\bar{n}_1} \cdots $, the polynomial $P$ has zero $\mathfrak{N}$-degree provided $\sum_A m_A + \sum_M \bar{n}_M = \sum_A n_A + \sum_M \bar{m}_M$:
\bea
\mathfrak{N}(P) &= &0  \; \;  \; \hbox{ with } P = (\xi^1)^{m_1} (\xi^2)^{m_2} \cdots (\eta^1)^{\bar{m}_1} \cdots (\xi_1)^{n_1} \cdots (\eta_1)^{\bar{n}_1} \cdots \nonumber \\
&\Leftrightarrow&  \sum_A m_A + \sum_M \bar{n}_M = \sum_A n_A + \sum_M \bar{m}_M.
\eea
We see that the only quadratic monomials that have zero $\mathfrak{N}$-degree are the $(N+2)^2$ monomials $\xi_A \eta_M$, $\xi^A \xi_B$, $\eta^M \eta_N$ and  $\xi^A \eta^M$.  The monomials $\xi^A \xi^B$,  $\xi^A \eta_M$, $\eta_M \eta_N$ have $\mathfrak{N}$-degree equal to 2 while the monomials $\xi_A \xi_B$,  $\xi_A \eta^M$, $\eta^M \eta^N$ have $\mathfrak{N}$-degree equal to -2. Let us introduce the function
\be
\mathcal{N} = \sum_A \xi^A \xi_A - \sum_M \eta^M \eta_M \ . 
\ee
For a given monomial $P$, we find that
\be
[\mathcal{N}, P]_\star = \mathfrak{N}(P) \, P \ . 
\label{UsefulRelationDegree}
\ee
Therefore, a function has zero $\mathfrak{N}$-degree if and only if it $\star$-commutes with the function $\mathcal{N}$.

Having prepared for two sets of unitary superalgebras and two super-variables associated with each,
the u$(1,1|N)$ superalgebra is then realizable in terms of the bilinears of the two super-variables that have zero $\mathfrak{N}$-degree
\bea
{\cal L}^- = \xi_A \eta_M, \qquad
{\cal L}^0 = \xi^A \xi_B \oplus \eta^M \eta_N, \quad
{\cal L}^+ = \xi^A \eta^M  \label{QuadraticBasisU}
\eea
and their $\star$-(anti)commutators\footnote{{For comparison with the literature using oscillator operators, we stress that the variables $z$ are here classical variables commuting or anticommuting exactly. The functions $f(z)$  are  the ``symbols" of the operators, with a quantization prescription - here Weyl symmetric ordering - that leads to the above star bracket. There is no ordering problem at the level of the symbols (for instance $q_1 q^1 = q^1 q_1$ even though $q_1 \star q^1 \not = q^1 \star q_1$). This is in contrast with the literature that we have quoted, for which the variables are quantum operators.}}. Stated differently, under the $\star$-algebra (\ref{su-starproduct}), the graded quadratic polynomials (binomials) of vanishing $\mathfrak{N}$-degree in the variables $q_i, \psi_\alpha$ and their complex conjugates form an algebra isomorphic to u$(1,1|N)$.

To reduce this further to su$(1,1 \vert N)$, one observes that $\mathcal{N}$ itself generates the u(1) subalgebra inside u$(1,1|N)$.  It is in fact, by construction,  in the center of u$(1,1|N)$.  Therefore, the superalgebra su$(1,1 \vert N)$ is obtained by taking the quotient of u$(1,1|N)$ by its center\footnote{The $N=2$ case, su$(1,1|N)$, is exceptional as the two bosonic subalgebras have same rank. Therefore, the supertrace of the unit supermatrix vanishes,  hence the diagonal u(1) in (\ref{maximalsub}) is absent. Nevertheless, the above construction works essentially in the same way.}.

To construct the corresponding infinite-dimensional higher-spin superalgebra ${\cal A}$, we consider the polynomials of all even degrees that have zero $\mathfrak{N}$-degree and which have no constant term. We also define the algebra in terms of the star product and quotient out the quadratic polynomials proportional to $\mathcal{N}$.  Note that an arbitrary polynomial of even polynomial degree and of zero $\mathfrak{N}$-degree is just a polynomial in the quadratic monomials appearing in (\ref{QuadraticBasisU}).

Evidently, the polynomials in ${\cal A}$ are closed among themselves under the $\star$-product {because the $\star$-(graded)commutator of two polynomials of zero $\mathfrak{N}$-degree has itself zero $\mathfrak{N}$-degree.  This follows from (\ref{UsefulRelationDegree}) and the Jacobi identity. The resulting algebra ${\cal A}$ is an infinite-dimensional superalgebra that contains {exactly} $\textrm{su}(1,1|N)$ as the sub-superalgebra of quadratic polynomials. It has an infinite number of generators of increasing spin, formed by the graded monomials of the $(N+2)$-variables of even degree. An element of ${\cal A}$ takes the form

\be
\bal
f &= f^{\alpha\beta} q_\alpha q_\beta + f^{\alpha i} q_\alpha \psi_i + f^{ij}\psi_i\psi_j + f_{\alpha \beta} q^\alpha q^\beta + f_{\alpha i} q^\alpha \psi^i + f_{ij} \psi^i \psi^j\\
& + f^\alpha_\beta q_\alpha q^\beta + f^\alpha_i q_\alpha \psi^i + f_\alpha^i q^\alpha \psi_i + f^i_j \psi_i \psi^j \\
& + f^{\alpha\beta \gamma \delta}q_\alpha q_\beta q_\gamma q_\delta+ f^{\alpha\beta \gamma i}q_\alpha q_\beta q_\gamma \psi_i + f^{\alpha\beta ij}q_\alpha q_\beta \psi_i\psi_j + \dots \\
&+ f^{\alpha \beta \gamma}_\delta q_\alpha q_\beta q_\gamma q^\delta + f^{\alpha \beta i}_\gamma q_\alpha q_\beta q^\gamma \psi_i + f^{\alpha \beta \gamma}_i q_\alpha q_\beta q_\gamma \psi^i + \cdots \\
& \cdots \\
& + f_{\alpha \beta \gamma \delta} q^\alpha q^\beta q^\gamma q^\delta + f_{\alpha \beta \gamma i} q^\alpha q^\beta q^\gamma \psi^i + f_{\alpha \beta ij} q^\alpha q^\beta \psi^i \psi^j + \cdots \\
& + \dots .
\eal
\label{su-expansion}
\ee
{where  $f^{\alpha \beta \cdots}$ etc. are complex-valued coefficients restricted by the vanishing $\mathfrak{N}$-degree condition. For instance, $f_{11}= f_{22}= f^{11} = f^{22} = f^{1}_{2} = f^{2}_1 = 0$ since the terms $(q^1)^2$, $(q_2)^2$, $q^1q_2$ which have $\mathfrak{N}$-degree $2$ are forbidden (and similarly for $(q_1)^2$, $(q^2)^2$, $q_1 q^2$ which have $\mathfrak{N}$-degree $-2$,  etc.)}

With the complex conjugation, these polynomials may be further decomposed into real and imaginary parts. Under the $\star$-product (\ref{su-starproduct}), only the imaginary parts of the polynomials are closed among themselves.
We thus consider the algebra ${\cal A}^{\rm E}$ formed by the anti-Hermitian (pure imaginary) polynomials of even degree under the $\star$-product. Under this reality condition, the coefficients of (\ref{su-expansion}) are related by complex conjugation, which swaps the covariant indices with the contravariant indices and vice versa:
\bea
\left(f^{\alpha \beta \cdots i j \cdots }_{ \gamma \delta \cdots k l \cdots}\right)^*
= - f_{\alpha \beta \cdots i j \cdots}^{\gamma \delta \cdots k l \cdots}, \quad \mbox{etc.}
\eea
This way, we have obtained the infinite-dimensional Lie superalgebra shs$(N|1,1)$ as the space of elements of ${\cal A}^{\rm E}$ equipped with the $\star$-product (\ref{su-starproduct}).

The higher-spin AdS$_3$ supergravity based on shs$(N|1,1)$ is then formulated as the Chern-Simons super-gauge theory with gauge algebra shs$(N|1,1)$. Expanded in terms of higher-spin generators, the super-connection takes the form of (\ref{su-expansion}). The reality condition of the polynomial now asserts that the super-connection is Hermitian.

It is helpful to recapitulate the relation of this superalgebra to the shs$^{\rm E}(N|2, \mathbb{R})$ superalgebra within the oscillator construction. In section 2, we constructed shs$^{\rm E}(N|2, \mathbb{R})$  upon the superalgebra osp$(N|2, \mathbb{R})$. This superalgebra can be constructed also from the unitary superalgebra. Whereas the shs$^{\rm E}(N|1,1)$ superalgebra above was built out of two sets of unitary superalgebras and two super-variables, the shs$^{\rm E}(N|2 \mathbb{R})$ superalgebra is constructible from just one set of unitary superalgebra su$(1|N)$ and two super-variables $\xi_A, \eta_B$ associated with it. The generators of su$(1|N)$ are the quadratic monomials
\bea
{\cal L}_0 = \xi_A \xi^B + \eta_A \eta^B.
\eea

If we also add
\bea
&& {\cal L}_- = \xi_{A} \eta_{B} + \eta_A \xi_B \\
&& {\cal L}_+ = \xi^A \eta^B + \eta^A \xi^B,
\label{symm}
\eea
the superalgebra is enlarged to osp$(2N|2, \mathbb{R})$. The only difference between this construction and that of section 2 is that the super-variables $\xi_A, \eta_B$ are non-Hermitian, whereas $q, \psi$ in section 2 are Hermitian.
This is purely a matter of convention, and the rest of the construction of the polynomials of infinite-degree and resulting infinite-dimensional higher-spin superalgebra follows essentially the same steps.  

With little work, another infinite-family class of higher-spin superalgebras shs$(2^*\vert N)$ can be constructed starting from the finite-dimensional superalgebras osp$(4^*|2N)$, whose representations are studied in \cite{Gunaydin:1990ag}. These superalgebras have bosonic subalgebras o$(4)^* \oplus$ usp$(2N)$, where o$(4)^*$ is isomorphic to su$(2) \oplus$ su$(1,1)$ and has as maximal compact subalgebra u$(2)$.
Thus, the osp$(4^* | 2N)$ superalgebra can also be realized in terms of the unitary superalgebra u$(2|N)$ and two sets of super-variables $\xi_A, \eta_B$ each of which consist of 2 bosonic and $N$ fermionic variables.

The superalgebra is then constructed from
the quadratic monomials of the super-variables
\bea
&& {\cal L}_0 = \xi_A \xi^B + \eta_A \eta^B \\
&& {\cal L}_- = \xi_A \eta_B - \eta_A \xi_B \\
&& {\cal L}_+ = \xi^A \eta^B - \eta^A \xi^B.
\eea
Here again, the complex conjugated variables are denoted as $(\xi_A)^\dagger = \xi^A, (\eta_A)^\dagger = \eta^A$. Compared to (\ref{symm}), here one chooses antisymmetric quadratic polynomials for ${\cal L}_\pm$. Apart from this, all other steps proceed essentially in the same way. By extending to the polynomials of arbitrary degree, one obtains an algebra whose real-form gives rise to a new higher-spin superalgebra shs$(2^*\vert N)$. 

Exceptional higher-spin superalgebras based on the superalgebras $F(4), G(3), D^1(2,1;\alpha)$ are constructible analogously. Since these constructions are not the main focus of the present paper, we shall relegate details of all these constructions to a separate publication.

\end{document}